\documentclass[a4paper,twoside]{article}

\usepackage{amsmath}
\usepackage{rotating}
\usepackage{color}
\usepackage{enumitem}

\newcommand{\affil}[1]{$^{\rm #1}$}
\newcommand{\be}{\begin{equation}}
\newcommand{\beq}{\begin{equation}}
\newcommand{\ba}{\begin{eqnarray}}
\newcommand{\ee}{\end{equation}}
\newcommand{\eeq}{\end{equation}}
\newcommand{\ea}{\end{eqnarray}}

\newcommand{\hs}{\hspace{1mm}}

\newcommand{\apj}{ApJ}
\newcommand{\na}{NA}
\newcommand{\aap}{A\&A}
\newcommand{\apjl}{ApJL}
\newcommand{\mnras}{MNRAS}
\newcommand{\aj}{AJ}
\newcommand{\sd}{\sigma_{\rm dust}}
\newcommand\ion[2]{#1$\;${\scshape{#2}}}
\newcommand{\fesc}{f^{\rm ion}_{\rm esc}}
\newcommand{\fesca}{f^{\alpha}_{\rm esc}}
\newcommand{\fesce}{f^{\rm eff}_{\rm esc}}
\newcommand{\tigm}{\mathcal{T}_{\rm IGM}}
\newcommand{\xo}{x_{\rm out}}
\newcommand{\xin}{x_{\rm in}}
\newcommand{\av}{a_{\rm v}}

\newcommand{\apjs}{ApJS}
\newcommand{\procspie}{PROCSPIE}
\newcommand{\nat}{{\it Nature}}
\newcommand{\araa}{ARA\&A}
\newcommand{\pasj}{PASJ}
\newcommand{\physrep}{Physics Reports}

\newcommand{\taud}{\tau_{\rm D}}
\newcommand{\tauHII}{\tau_{\rm HII}}

\usepackage[authoryear]{natbib}
\usepackage{epsfig}

\bibpunct{(}{)}{;}{a}{}{,}
\date{} %Please leave the date blank
\def\lsim{~\rlap{$<$}{\lower 1.0ex\hbox{$\sim$}}}

\def\gsim{~\rlap{$>$}{\lower 1.0ex\hbox{$\sim$}}}
\title{\large\bf\flushleft Ly$\alpha$ Emitting Galaxies as a Probe of Reionization}
\author{\parbox{\textwidth}{\flushleft
\vspace{-0.5cm}
{\it Mark Dijkstra\affil{A,}\affil{B}}\\
\vspace{0.4cm}
{\small \affil{A}\, Institute of Theoretical Astrophysics, University of Oslo, Postboks 1029, 0858 Oslo, Norway, mark.dijkstra@astro.uio.no}\\
{\small \affil{B}\, MPI fuer Astrophysik, Karl-Schwarzschild-Str. 1, 85741 Garching, Germany}}}
\begin{document}
\twocolumn[
\begin{changemargin}{.8cm}{.5cm}
\begin{minipage}{.9\textwidth}
\vspace{-1cm}
\maketitle
%
%
%%%%%%%%%%%%%     ABSTRACT    %%%%%%%%%%%%%
\small{\bf Abstract:}
The Epoch of Reionization (EoR) represents a milestone in the evolution of our Universe. Star-forming galaxies that existed during the EoR likely emitted a significant fraction ($\sim 5-40\%$) of their bolometric luminosity as Ly$\alpha$ line emission. However, neutral intergalactic gas that existed during the EoR was opaque to Ly$\alpha$ emission that escaped from galaxies during this epoch, which makes it difficult to observe. The neutral intergalactic medium (IGM) may thus reveal itself by suppressing the Ly$\alpha$ flux from background galaxies. 
Interestingly, a `sudden' reduction in the observed Ly$\alpha$ flux has now been observed in galaxies at $z >6$. This review contains a detailed summary of Ly$\alpha$ radiative processes: I describe ({\it i}) the main Ly$\alpha$ emission processes, including collisional-excitation \& recombination (and derive the origin of the famous factor `0.68'), and ({\it ii}) basic radiative transfer concepts, including e.g. partially coherent scattering, frequency diffusion, resonant versus wing scattering, optically thick versus 'extremely' optically thick (static/outflowing/collapsing) media, and multiphase media. Following this review, I derive expressions for the Gunn-Peterson optical depth of the IGM during (inhomogeneous) reionization and post-reionization. I then describe why current observations appear to require a very rapid evolution of volume-averaged neutral fraction of hydrogen in the context of realistic inhomogeneous reionization models, and discuss uncertainties in this interpretation. Finally, I describe how existing \& futures surveys and instruments can help reduce these uncertainties, and allow us to fully exploit Ly$\alpha$ emitting galaxies as a probe of the EoR. 

%%%%%%%%%%%%%     KEYWORDS    %%%%%%%%%%%%%
\medskip{\bf Keywords:} cosmology: dark ages, reionization, first stars --- galaxies: intergalactic medium, high redshift --- radiative transfer --- scattering --- ultraviolet: galaxies
% Please select up to six keywords, written in lower case. PASA uses the
% standard list of subject headings adopted by The Astrophysical Journal
% and available from http://aas.org/authors/astronomical-subject-keywords-update-october-2009 .
% Keywords are separated by em-dashes, i.e. ---

%%%%%%%%DO NOT EDIT%%%%%%%%%%%%
\medskip
\medskip
\end{minipage}
\end{changemargin}
]
\small
%%%%%%%%EDIT FROM HERE%%%%%%%%%%%%

\section{Introduction}
%Please see the PASA Style Guide for help with correct layout for your manuscript.
%Examples of tables and figures are given below.

\begin{table*}[ht]
\begin{center}
\caption{Summary of symbols used throughout this paper.}\label{table:symbols}
\scriptsize{\begin{tabular}{lll}
\hline \hline 
  Symbol & meaning & comment \\
  \hline
\multicolumn{3}{|l|}{Fundamental physical constants.}\\ 
\hline
$e$ & electron/proton charge & $e=4.80\times 10^{-10}$ esu \\
$h_P$ &Planck's constant  & $h_P=6.63\times 10^{-27}$ erg s \\
$k_{\rm B}$ & Boltzmann's constant& $k_{\rm B}=1.38 \times 10^{-16}$ erg/K\\
$c$ & speed of light & $c=3.00\times 10^{10}$ cm s$^{-1}$\\
$m_p$ & proton mass& $m_p=1.67\times 10^{-24}$ gram\\
$a_0$& Bohr radius & $a_0=5.29\times 10^{-11}$ cm\\
\hline
\multicolumn{3}{|l|}{Parameters in emission processes.}\\ 
\hline
$n_{\rm HI}$ & number density of hydrogen atoms & \\
$n_e$ & number density of electrons & \\
$n_p$& number density of protons & \\
$\alpha_{\rm A/B}$ & case A/B recombination coefficient & \\
& case-B/A: optically thick/thin to Lyman-series lines & \\
$\alpha_{\rm nl}$ & state-specific recombination coefficient& \\
$A_{\rm n,l,n',l'}$ &Einstein-A coefficient for transition $(n,l)\rightarrow (n',l')$& \\
$A_{\alpha}$ &Einstein-A coefficient for the Ly$\alpha$ transition 2p$\rightarrow$1s&$A_{\alpha}=6.25\times 10^8$ s$^{-1}$ \\
\hline
\multicolumn{3}{|l|}{Parameters in Radiative Transfer.}\\ 
\hline
$E_{\alpha}$ & energy of Ly$\alpha$ photon& $E_{\alpha}=10.2$ eV \\
$\lambda_{\alpha}$ & wavelength of the Ly$\alpha$ transition& $\lambda_{\alpha}=1215.67$ \AA \\
$v_{\rm th}$ & `thermal' velocity of \ion{H}{I} in cold clumps & $v_{\rm th}=\sqrt{2kT/m_{\rm p}}=12.9(T/10^{4}{\rm  K})^{1/2}$ km/s\\
$\nu_{\alpha}$ & Ly$\alpha$ resonance frequency & $\nu_{\alpha}=2.46 \times 10^{15}$ Hz\\
$\Delta \nu_{\rm D}$ & thermal line broadening  & $\Delta \nu_{\rm D} \equiv \nu_{\alpha}v_{\rm th}/c$\\ 
$\nu$ & photon frequency & \\ 
$x$ & dimensionless photon frequency & $x\equiv (\nu-\nu_{\alpha})/\Delta \nu_{\alpha}$\\ 
$\sigma_0$ & Ly$\alpha$ absorption cross section at line center & $\sigma_0=5.88 \times 10^{-14}(T/10^{4}{\rm  K})^{-1/2}$ cm$^{2}$\\ 
$\tau_0$ & line centre optical depth & $\tau_0 = N_{\rm HI} \sigma_0$\\
$\sigma_{\alpha}(x/\nu)$ & Ly$\alpha$ absorption cross section at frequency $x$/$\nu$ & $\sigma_{\alpha}(x)=\sigma_0\phi(x)$\\
$\phi(x)$ & Voigt function at frequency $x$ &  We adopt $\phi(x=0)=1$, i.e.\\
& & $\int \phi(x)dx=\sqrt{\pi}$ and $\int \phi(\nu)d\nu=\sqrt{\pi} \Delta \nu_{\rm D}$\\ 
$\sd$ & dust absorption cross section per hydrogen atom & \\
$A_{\rm dust}$ & albedo of dust, denotes probability that & \\
& dust grain {\it scatters} rather than destroys Ly$\alpha$ photon &\\
$\av$ & Voigt parameter & $\av=A_{\alpha}/[4\pi\Delta \nu_{\rm D}]=4.7 \times 10^{-4}(T/10^{4}{\rm K})^{-1/2}$ \\
${\bf k}_{\rm  in/out}$ & Unit vector that denotes the propagation direction & \\
& of the photons before/after scattering. &\\ 
$x_{\rm  in/out}$ & dimensionless photon frequency before/after scattering. &\\ 
$R(x_{\rm out},x_{\rm in})$ & angle averaged frequency redistribution function: & \\
& denotes the probability of having $x_{\rm out}$ given $x_{\rm in}$ & \\
$R(x_{\rm out},x_{\rm in},{\bf k}_{\rm out},{\bf k}_{\rm in})$ & directional dependent frequency redistribution function: & \\
& denotes probability of having $x_{\rm out}$,${\bf k}_{\rm out}$ given $x_{\rm in}$, ${\bf k}_{\rm in}$& \\
$\mu$ & Cosine of the scattering angle &$\mu ={\bf k}_{\rm in} \cdot {\bf k}_{\rm out}$\\ 
$P(\mu)$ &Scattering phase function & We normalize $\int _{-1}^{1}d\mu\hs P(\mu)=4 \pi$\\  
\hline
\multicolumn{3}{|l|}{Parameters Relevant for Describing Ly$\alpha$ Emitting Galaxies.}\\ 
\hline
$L_{\alpha}$ & Ly$\alpha$ luminosity  & \\
$\fesc$ & escape fraction of ionising photons& \\
$\fesca$ & real escape fraction of Ly$\alpha$ photons & \\
$\tigm$ & fraction of Ly$\alpha$ photons transmitted through the IGM & \\
$\fesce$ & `effective' escape fraction of Ly$\alpha$ photons& \\
$\taud$ & optical depth in neutral patches of intergalactic gas & \\
$\tauHII$ & opacity of IGM in ionized gas/bubbles & \\
$\tau_{\rm IGM}$ &  total opacity of IGM & $\tau_{\rm IGM}=\tauHII+\taud$ \\
\hline
\end{tabular}}
%\medskip\\
%$^a$Table footnotes go here.\\
\end{center}
%\label{table:symbols}
\end{table*}

The Epoch of Reionization (EoR) represents a milestone in the evolution of our Universe. It represents the last major phase transformation of its gas from a cold (i.e. a few tens of K) and neutral to a fully ionized, hot (i.e. $\approx 10^4$ K) state. This transformation was likely associated with the formation of the first stars, black holes and galaxies in our Universe. Understanding the reionization process is therefore intimately linked to understanding the formation of the first structures in our Universe -- which represents one of the most basic and fundamental questions in astrophysics. 

Reionization is still not well constrained: anisotropies in the Cosmic Microwave Background (CMB) constrain the total optical depth to scattering by free electrons\footnote{Inhomogeneous reionization further affects the CMB anistropies on arcminute scales via the kinetic SZ (kSZ) effect. However, current measurement of the power-spectrum on these scales do not allow for stringent constraints \citep[see][for a discussion]{Mesinger12}. } to be $\tau_{\rm e}=0.088 \pm 0.013$ \citep[][]{Komatsu11,Hinshaw13,Planck}. For a reionization history in which the Universe transitions from fully neutral to fully ionized at redshift $z_{\rm reion}$ over a redshift range $\Delta z=0.5$, this translates to $z_{\rm reion}=11.1 \pm 1.1$. Gunn-Peterson troughs that have been detected in the spectra of quasars $z \gsim 6$ \citep[e.g.][]{Becker01,Fan,Mortlock,Venemans} suggest that the intergalactic medium (IGM) contained a significant neutral fraction \citep[with a volume averaged fraction $\langle x_{\rm HI} \rangle_{\rm V}\sim 0.1$, see e.g.][]{WL03,Mesinger07,Bolton11,Schroeder}. Moreover, quasars likely inhabit highly biased, overdense regions of our Universe, which probably were reionized earlier than the Universe as a whole. It has been shown that existing quasar spectra at $z>5$ are consistent with a significant neutral fraction, $\langle x_{\rm HI} \rangle_{\rm V}\sim 0.1$ even at  $z\sim 5$ \citep[][]{Mesinger10,McGreer}.  

The constraints obtained from quasars and the CMB thus suggest that reionization was a temporally extended process that ended at $z\sim 5-6$, but that likely started at $z \gg 11$ \citep[e.g.][]{Pritchard10,Mitra}. These constraints are consistent with those obtained measurements of the temperature of the IGM at $z>6$ \citep[][]{Theuns02,HH03,Raskutti}, observations of the Ly$\alpha$ damping wing in gamma-ray burst after-glow spectra \citep{Totani,McQuinn08}, and Ly$\alpha$ emitting galaxies \citep[e.g.][]{Haiman99,MR04,K06}. \\

In this review I will discuss why Ly$\alpha$ emitting galaxies provide a unique probe of the EoR, and place particular emphasis on describing the physics of the relevant Ly$\alpha$ radiative processes. Throughout, `Ly$\alpha$ emitting galaxies' refer to all galaxies with `strong' Ly$\alpha$ emission (what `strong' means is clarified later), and thus includes both LAEs (Ly$\alpha$ emitters) and Ly$\alpha$ emitting drop-out galaxies. We refer to LAEs as Ly$\alpha$ emitting galaxies that have been {\it selected} on the basis of their Ly$\alpha$ line. This selection can be done either in a spectroscopic or in a narrow-band (NB) survey. NB surveys apply a set of color-color criteria that define LAEs. This typically requires some excess flux in the narrow-band which translates to a minimum EW$_{\rm min}$ of the line. It is also common in the literature to use the term LAE to refer to {\it all} galaxies for which the Ly$\alpha$ EW$>$EW$_{\rm min}$ (irrespective of how these were selected).The outline of this review is as follows: In \S~\ref{sec:RT} I give the general radiative transfer equation that is relevant for Ly$\alpha$. The following sections contain detailed descriptions of the components in this equation:

\begin{itemize}[leftmargin=*]

\item In \S~\ref{sec:emission} I summarize the main Ly$\alpha$ emission processes, including collisional-excitation \& recombination. For the latter, I derive the origin of the factor $'0.68'$ (which denotes the number of Ly$\alpha$ photons emitted per recombination event), which is routinely associated with case-B recombination. I will also describe why and where departures from case-B may arise, which underlines why star-forming galaxies are thought to have very strong Ly$\alpha$ emission lines, especially during the EoR (\S~\ref{sec:emission}).

\item In \S~\ref{sec:transfer} I describe the basic radiative transfer concepts that are relevant for understanding Ly$\alpha$ transfer. These include for example, partially coherent scattering, frequency diffusion, resonant versus wing scattering, and optically thick versus 'extremely' optically thick in static/ outflowing/ collapsing media. 

\end{itemize}

After this review, I discuss our current understanding of Ly$\alpha$ transfer at interstellar and intergalactic level in \S~\ref{sec:realRT}. With this knowledge, I will then discuss the impact of a neutral intergalactic medium on the visibility of the Ly$\alpha$ emission line from galaxies during the EoR (\S~\ref{sec:EoR}). I will then apply this to existing observations of Ly$\alpha$ emitting galaxies, and discuss their current constraints on the EoR. This discussion will show that existing constraints are still weak, mostly because of the limited number of known Ly$\alpha$ emitting galaxies at $z>6$. However, we expect the number of known Ly$\alpha$ emitting galaxies at $z>6$ to increase by up to two orders of magnitude. I will discuss how these observations (and other observations) are expected to provide strong constraints on the EoR within the next few years in \S~\ref{sec:outlook}. Table~\ref{table:symbols} provides a summary of symbols used throughout this review.

\section{Radiative Transfer Equation}
\label{sec:RT}

The change in the intensity of radiation $I(\nu)$ at frequency $\nu$ that is propagating into direction ${\bf n}$ (where $|{\bf n}|=1$) is given by \citep[e.g.][]{RL79}

\begin{align}
\label{eq:RT}
{\bf n}\cdot \nabla I(\nu,{\bf n})=-\alpha(\nu)I(\nu,{\bf n})+j(\nu,{\bf n}) \\ \nonumber
+ \int d\Omega' \int d {\bf n}' I(\nu',{\bf n}')R(\nu',\nu,{\bf n}',{\bf n}),
\end{align} where 

\begin{itemize}[leftmargin=*]

\item the attenuation coefficient $\alpha(\nu) \equiv n_{\rm HI} (\sigma_{\alpha}(\nu)+\sd)$, in which $\sigma_{\alpha}(\nu)$ denotes the Ly$\alpha$ absorption cross section, and $\sd$ denotes the dust absorption cross section {\it per hydrogen nucleus}. I give an expression for $\sigma_{\alpha}(\nu)$ in \S~\ref{sec:cross}, and for $\sd$ in \S~\ref{sec:dust}.

\item $j(\nu)$ denotes the volume emissivity (energy emitted per unit time, per unit volume) of Ly$\alpha$ photons, and can be decomposed into $j(\nu) = j_{\rm rec}(\nu) + j_{\rm coll}(\nu)$. Here, $j_{\rm rec}(\nu)$ /$j_{\rm coll}(\nu)$ denotes the contribution from recombination (see \S~\ref{sec:rec})/collisional-excitation (see \S~\ref{sec:co}).

\item $R(\nu',\nu,{\bf n}',{\bf n})$ denotes the `redistribution function', which measures the probability that a photon of frequency $\nu'$ propagating into direction ${\bf n}'$ is scattered into direction ${\bf n}$ and to frequency $\nu$. In \S~\ref{sec:redist} we discuss this redistribution function in more detail. 
\end{itemize}
Eq~\ref{eq:RT} is an integro-differential equation, and has been studied for decades \citep[e.g.][]{Chandra45,Unno50,Harrington73,Neufeld90,Yang11,HM12}. Eq~\ref{eq:RT} simplifies if we ignore the directional dependence of the Ly$\alpha$ radiation field (which is reasonable in gas that is optically thick to Ly$\alpha$ photons)\footnote{Ignoring the directional dependence of $I(\nu)$ allows us replace the term ${\bf n}\cdot \nabla \rightarrow \frac{d}{ds}$, and replace $I({\bf n},\nu)$ with the angle averaged intensity $J(\nu)\equiv \int d{\bf n} I(\nu,{\bf n})$.}, and the directional dependence of the redistribution function. \citet{R94} showed that the `Fokker-Planck' approximation - a Taylor expansion in the angle averaged intensity $J(\nu)$ in the integral - allows one to rewrite Eq~\ref{eq:RT} as a differential equation \citep[also see][]{HM12}:
\begin{eqnarray}
 \frac{dJ(\nu)}{d\tau}=\frac{(\Delta \nu_{\rm D})^2}{2}\frac{\partial}{\partial \nu} \phi(\nu) \frac{\partial J(\nu)}{\partial \nu},
\label{eq:RTdiff}
\end{eqnarray} where we replaced the attenuation coefficient [$\alpha(\nu)$] with the optical depth $d\tau(\nu) \equiv \alpha(\nu)ds$, in which $ds$ denotes a physical infinitesimal displacement. Furthermore, we set $j(\nu,{\bf n})=0$ and $\sd=0$ for simplicity. Eq~\ref{eq:RTdiff} is a diffusion equation. Ly$\alpha$ transfer through an optically thick medium is therefore a diffusion process: as photons propagate away from their source, they diffuse away from line centre. That is, the Ly$\alpha$ transfer process can be viewed as diffusion process in real and frequency space.

\section{Ly$\alpha$ Emission}
\label{sec:emission}
Unlike UV-continuum radiation, the majority of Ly$\alpha$ line emission typically does not originate in stellar atmospheres. Instead, Ly$\alpha$ line emission is predominantly powered via two other mechanisms. In the {\it first}, ionizing radiation emitted by hot young O and B stars ionize their surrounding, dense interstellar gas, which recombines on a short timescale,  $t_{\rm rec} = 1/n_{\rm e} \alpha \sim 10^5\hs{\rm yr}\hs (n_{\rm e}/1 \hs {\rm cm}^{-3})$ $(T/10^4\hs{\rm K})^{0.7}$ \citep[e.g.][]{HG97}. A significant fraction of the resulting recombination radiation emerges as Ly$\alpha$ line emission (see e.g. Johnson et al. 2009, Raiter et al. 2010, Pawlik et al. 2011 and \S~\ref{sec:rec}). 

In the {\it second}, Ly$\alpha$ photons are emitted by collisionally-excited HI. As we discuss briefly in \S~\ref{sec:co}, the collisionally-excited Ly$\alpha$ flux emitted by galaxies appears subdominant to the Ly$\alpha$ recombination radiation, but may become more important towards higher redshifts.

\subsection{Recombination Radiation: The Origin of the Factor `0.68'}
\label{sec:rec}

\begin{figure*}[t]
\begin{center}
\vbox{\centerline{\epsfig{file=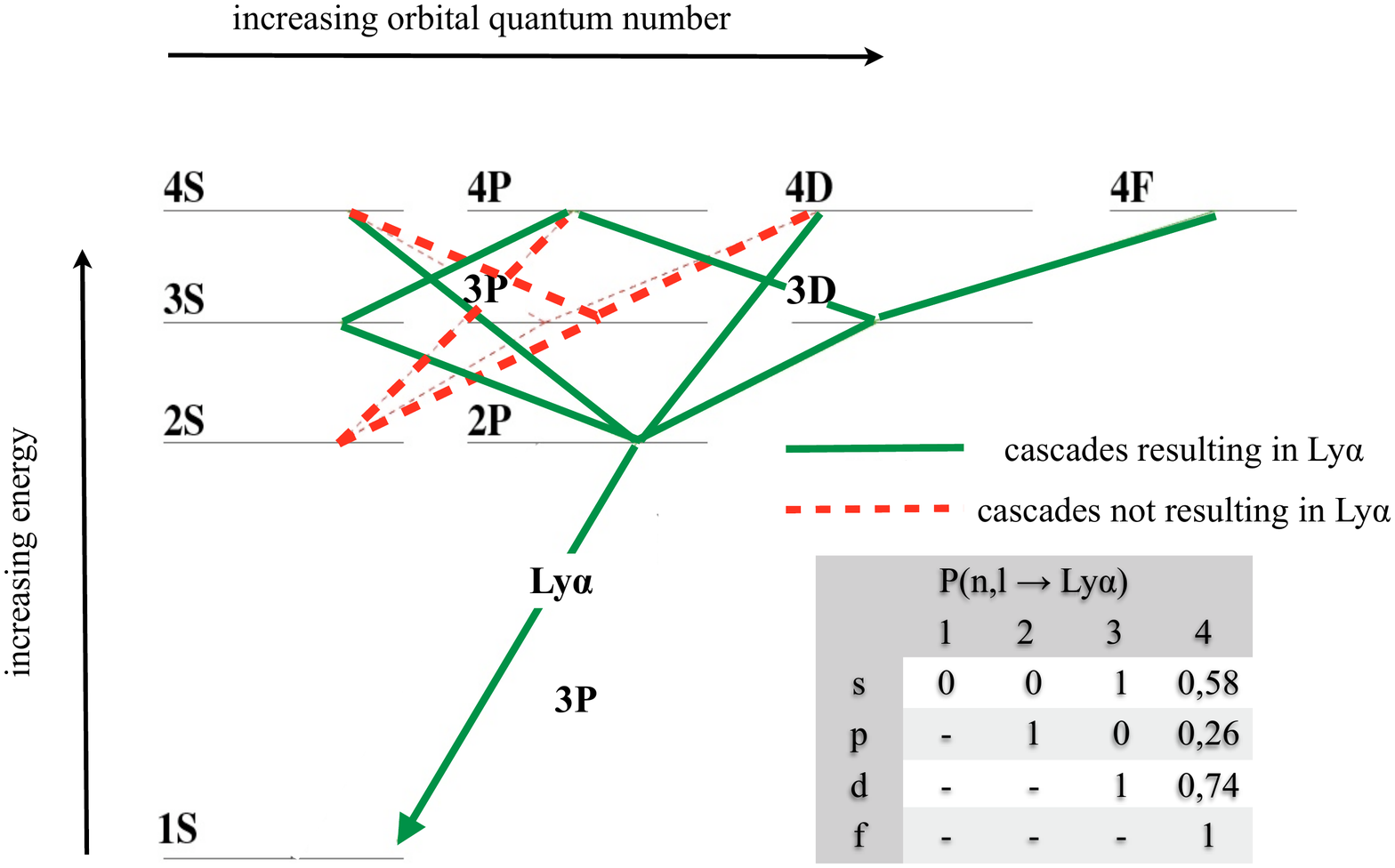,angle=0,width=14.0truecm}}}
\vspace{-5mm}
\caption{This Figure shows a schematic diagram of the energy levels of a hydrogen atom. The energy of a quantum state increases from bottom to top. Each state is characterized by two quantum numbers $n$ (principle quantum number) and $l$ (orbital quantum number). Recombination can put the atom in any state $nl$, which then undergoes a radiative cascade to the groundstate (1S). Quantum selection rules dictate that the only permitted transitions have $\Delta l=\pm 1$. These transitions are indicated in the Figure. {\it Green lines} [{\it red dotted lines}] show cascades that [{\it do not}] result in Ly$\alpha$. The {\it lower right panel} shows that probability that a cascade from state $nl$ results in Ly$\alpha$, $P(n,l \rightarrow {\rm Ly}\alpha)$ (Eq~\ref{eq:pnllya}).}
\label{fig:schemeHI}
\end{center}
\end{figure*}

\begin{figure*}[t]
\vbox{\centerline{\epsfig{file=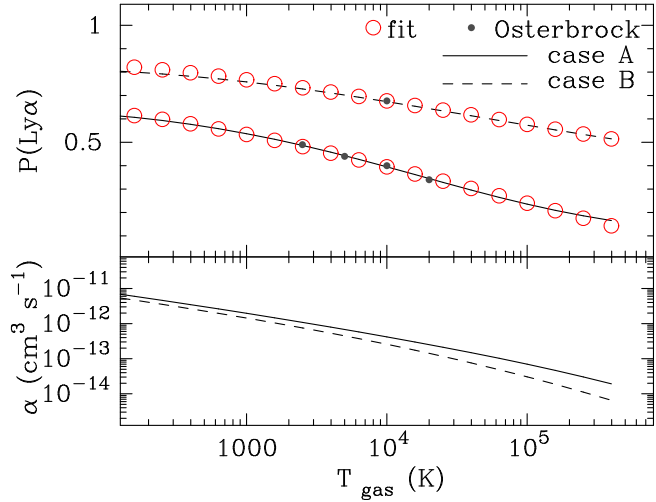,angle=0,width=10.5truecm}}}
\caption[]{The {\it top panel}  shows the number of Ly$\alpha$ photons per recombination event, $P({\rm Ly}\alpha)$, as a function of temperature for case-A ({\it solid line}) and case-B ({\it dashed line}). In both cases, $P({\rm Ly}\alpha)$ decreases with temperature. For comparison, the {\it filled circle} at $[T,P_{\rm B}({\rm Ly}\alpha)]=[10^4,0.68]$ is the number that is given by \citet{Osterbrock89} and which is commonly used in the literature (other values given in Osterbrock 1989 are shown as {\it filled circles}). The {\it top panel} shows for example that at $T=0.5\times 10^4$ K ($T=2\times 10^4$ K), $P_{\rm B}({\rm Ly}\alpha)=0.70$ ($P_{\rm B}({\rm Ly}\alpha)=0.64$).  A stronger temperature dependence is found for case-A recombination.  The {\it bottom panel} shows the total case-A and case-B recombination coefficients. The recombination coefficient $\alpha_B$ decreases more rapidly with temperature than $\alpha_A$, which implies that the fractional contribution from direct recombination into the ground state increases with temperature. Generally, as temperature increases a larger fraction of recombination events goes into the other low$-n$ states which reduces the number of Ly$\alpha$ photons per recombination event. {\it Red open circles} represent fitting formulae given in Eq~\ref{eq:fitrec}.}
\label{fig:resultrec}
\end{figure*} 

The volume Ly$\alpha$ emissivity following recombination is often given by
\begin{equation}
j_{{\rm rec}}(n,T\,\nu)=0.68 n_{\rm e} n_{\rm p} \alpha_{\rm B}(T)\phi(\nu)E_{\alpha}/(\sqrt{\pi}\Delta \nu_{\rm D}),
\end{equation} where $E_{\alpha}=10.2$ eV,  $n_{\rm e}$/$n_{\rm p}$ denotes the number density of free electrons/protons, and $\phi(\nu)$ denotes the Voigt profile (normalised to $\int d\nu \hs \phi(\nu)=\sqrt{\pi}\Delta \nu_{\rm D}$, in which $\Delta \nu_{\rm D}=1.1\times 10^{11}(T/10^4\hs{\rm K})^{1/2}$ Hz quantifies thermal broadening of the line). Expressions for $\phi(\nu)$ are given in \S~\ref{sec:cross}. The factor $0.68$ denotes the fraction of recombination events resulting in Ly$\alpha$ and is derived next.\\

The capture of an electron by a proton generally results in a hydrogen atom in an excited state $(n,l)$. %
Once an atom is in a quantum state $(n,l)$ it radiatively cascades to the ground state $n=1$, $l=0$ via intermediate states $(n_i,l_i)$. The probability that a radiative cascade from the state $(n,l)$ results in a Ly$\alpha$ photon is given by   

\begin{eqnarray}
P(n,l\rightarrow  {\rm Ly}\alpha)= \\ \nonumber = \sum_{n',l'}P(n,l\rightarrow n',l')P(n',l'\rightarrow {\rm Ly}\alpha).
\label{eq:pnllya}
\end{eqnarray} That is, the probability can be computed if one knows the probability that a radiative cascade from lower excitation states $(n',l')$ results in the emission of a Ly$\alpha$ photon, {\it and} the probabilities that the atom cascades into these lower excitation states, $P(n,l \rightarrow n',l')$. This latter probability is given by

\begin{equation}
P(n,l\rightarrow n',l')=\frac{A_{n,l,n',l'}}{\sum_{n'',l''}A_{n,l,n'',l''}},
\end{equation} in which $A_{n,l,n',l'}$ denotes the Einstein A-coefficient for the $nl \rightarrow n'l'$ transition\footnote{This coefficient is given by

\begin{equation}
A_{n,l,n',l'}=\frac{64\pi^4\nu_{ul}^3}{3h_Pc^3}\frac{{\rm max}(l',l)}{2l+1}e^2a_0^2[M(n,l,n',l')]^2,
\label{eq:Anlnl}
\end{equation} where fundamental quantities $e$, $c$, $h_{\rm P}$, and $a_0$ are given in Table~\ref{table:symbols}, $h_{\rm P}\nu_{\rm ul}$ denotes the energy difference between the upper (n,l) and lower (n',l') state. The matrix $M(n,l,n',l')$ involves an overlap integral that involves the radial wavefunctions of the states $(n,l)$ and $(n',l')$:

\begin{equation}
M(n,l,n',l')=\int_0^{\infty}P_{n,l}(r)r^3P_{n',l'}(r)dr.
\end{equation} Analytic expressions for the matrix $M(n,l,n',l')$ that contain hypergeometric functions were derived by \citet{Gordon29}. For the Ly$\alpha$ transition $M(n,l,n',l')=\sqrt{6}(128/243)$ \citep{HB90}.}.

The quantum mechanical selection rules only permit transitions for which $|l-l'|=1$, which restricts the total number of allowed radiative cascades. Figure~\ref{fig:schemeHI} schematically depicts permitted radiative cascades in a four-level H atom. {\it Green solid lines} depict radiative cascades that result in a Ly$\alpha$ photon, while {\it red dotted lines} depict radiative cascades that do not yield a Ly$\alpha$ photon. 

Figure~\ref{fig:schemeHI} also contains a table that shows the probability $P(n',l'\rightarrow {\rm Ly}\alpha)$ for $n\leq 5$. For example, the probability that a a radiative cascade from the $(n,l)=(3,1)$ state (i.e. the 3p state) produces a Ly$\alpha$ photon is $0$, because the selection rules only permit the transitions $(3,1)\rightarrow(2,0)$ and $(3,1)\rightarrow(1,0)$. The first transition leaves the H-atom in the 2s state, from which it can only transition to the ground state by emitting two photons \citep{BT40}.  On the other hand, a radiative cascade from the $(n,l)=(3,2)$ state (i.e. the 3d state) will certainly produce a Ly$\alpha$ photon, since the only permitted cascade is $(3,2)\rightarrow(2,1)\overset{{\rm Ly}\alpha}{\rightarrow}(1,0)$.  Similarly, the only permitted cascade from the 3s state is $(3,0)\rightarrow(2,1)\overset{{\rm Ly}\alpha}{\rightarrow}(1,0)$ , and $P(3,0 \rightarrow {\rm Ly}\alpha)=1$. For $n>3$, multiple radiative cascades down to the ground state are generally possible, and $P(n,l\rightarrow {\rm Ly}\alpha)$ takes on values other than $0$ or $1$ \citep[see e.g.][]{SG51}.
 
The probability that an arbitrary recombination event results in a Ly$\alpha$ photon is given by

\begin{equation}
P({\rm Ly}\alpha)= \sum_{n_{\rm min}}^{\infty} \sum_{l=0}^{n-1}\frac{\alpha_{nl}(T)}{\alpha_{\rm tot}(T)}P(n,l\rightarrow {\rm Ly}\alpha)\,
\label{eq:plya}
\end{equation} where the first term denotes the fraction of recombination events into the $(n,l)$ state, in which $\alpha_{\rm tot}$ denotes the total recombination coefficient $\alpha_{\rm tot}(T)=\sum_{n_{\rm min}}^{\infty} \sum_{l=0}^{n-1}\alpha_{nl}(T)$.  The temperature-dependent state specific recombination coefficients $\alpha_{nl}(T)$ can be found in for example \citep{Burgess65} and \citet{Rubino06}. The value of $n_{\rm min}$ depends on the physical conditions of the medium in which recombination takes place, and two cases bracket the range of scenarios commonly encountered in astrophysical plasmas: 
\begin{itemize}[leftmargin=*]

\item {\it `case-A}' recombination: recombination takes place in a medium that is optically thin at all photon frequencies. In this case, direct recombination to the ground state is allowed and $n_{\rm min}=1$.

\item {\it `case-B}' recombination: recombination takes place in a medium that is opaque to all Lyman series\footnote{At gas densities that are relevant in most astrophysical plasmas, hydrogen atoms predominantly populate their electronic ground state ($n=1$), and the opacity in the Balmer lines is generally negligible.  In theory one can introduce {\it case-C/D/E/...} recombination to describe recombination in a medium that is optically thick to Balmer/Paschen/Bracket/... series photons.} photons (i.e. Ly$\alpha$, Ly$\beta$, Ly$\gamma$, ...), and to ionizing photons that were emitted following direct recombination into the ground state. In the so-called `on the spot approximation', direct recombination to the ground state produces an ionizing photon that is immediately absorbed by a nearby neutral H atom. Similarly, any Lyman series photon is immediately absorbed by a neighbouring H atom. This case is quantitatively described by setting $n_{\rm min}=2$, and by setting the Einstein coefficient for all Lyman series transitions to zero, i.e. $A_{np,1s}=0$.
 \end{itemize}

Figure~\ref{fig:resultrec} shows the total probability $P({\rm Ly}\alpha)$ (Eq~\ref{eq:plya}) that a Ly$\alpha$ photon is emitted per recombination event as a function of gas temperature $T$, assuming case-A recombination ({\it solid line}), and case-B recombination ({\it dashed line}). For gas at $T=10^4$ K and case-B recombination, we have $P({\rm Ly}\alpha)=0.68$. This value $`0.68'$ is often encountered during discussions on Ly$\alpha$ emitting galaxies. It is worth keeping in mind that the probability $P({\rm Ly}\alpha)$ increases with decreasing gas temperature and can be as high as $P({\rm Ly}\alpha)=0.77$ for $T=10^3$ K (also see Cantalupo et al. 2008). The {\it red open circles} represent the following two fitting formulae
\begin{align}
\label{eq:fitrec}
P_{\rm A}({\rm Ly}\alpha)=0.41-0.165 \log T_4-0.015(T_4)^{-0.44}\\ \nonumber
P_{\rm B}({\rm Ly}\alpha)=0.686-0.106 \log T_4-0.009(T_4)^{-0.44},
\end{align} where $T_4 \equiv T/10^4$ K. The fitting formula for case-B is taken from Cantalupo et al. (2008).\\

Recombinations in HII regions in the ISM are balanced by photoionization in equilibrium HII regions. The total recombination rate in an equilibrium HII region therefore equals the total photoionization rate, or the total rate at which ionizing photons are absorbed in the HII region (in an expanding HII region, the total recombination rate is less than the total rate at which ionising photons are absorbed). If a fraction $\fesc$ of ionizing photons is {\it not} absorbed in the HII region (and hence escapes), then the total Ly$\alpha$ production rate in recombinations can be written as

\begin{align}
\label{eq:0.68}
\dot{N}^{\rm rec}_{{\rm Ly}\alpha}=P({\rm Ly}\alpha)(1-f_{\rm esc})\dot{N}_{\rm ion} \approx \\ \nonumber
\approx  0.68(1-\fesc)\dot{N}_{\rm ion},\hs\hs{\rm {\it case}}-B,\hs{\rm T}=10^4\hs{\rm K}
\end{align} where $\dot{N}_{\rm ion}$ ($\dot{N}^{\rm rec}_{{\rm Ly}\alpha}$) denotes the rate at which ionizing (Ly$\alpha$ recombination) photons are emitted. The equation on the second line is commonly adopted in the literature. The ionizing emissivity of star-forming galaxies is expected to be boosted during the EoR: stellar evolution models combined with stellar atmosphere models show that the effective temperature of stars of fixed mass become hotter with decreasing gas metallicity \citep{TS00,Schaerer02}. The increased effective temperature of stars causes a larger fraction of their bolometric luminosity to be emitted as ionizing radiation. We therefore expect galaxies that formed stars from metal poor (or even metal free) gas during the EoR, to be strong sources of nebular emission. \citet{Schaerer03} provides the following fitting formula for $\dot{N}_{\rm ion}$ as a function of absolute gas metallicity\footnote{It is useful to recall that solar metallicity $Z_{\odot}=0.02$.} $Z_{\rm gas}$, $\log \dot{N}_{\rm ion}=-0.0029\times(\log Z_{\rm gas} +9.0)+53.81$,which is valid for a Salpeter IMF in the mass range $M=1-100M_{\odot}$.\\

A useful measure for the `strength' of the Ly$\alpha$ line (other than just its flux) is given by the equivalent width 
\begin{equation}
{\rm EW}\equiv \int d \lambda \hs (F(\lambda) - F_0)/F_0,
\end{equation} which measures the total line flux compared to the continuum flux density just redward (as the blue side can be affected by intergalactic scattering, see \S~\ref{sec:IGM}) of the Ly$\alpha$ line, $F_0$. For `regular' star-forming galaxies (Salpeter IMF, solar metallicity) the maximum physically allowed restframe EW is EW$_{\rm max}\sim 240$ \AA\hs \citep[see e.g.][and references therein]{Schaerer03,Laursen13}. Reducing the gas metallicity by as much as two orders of magnitude typically boosts the EW$_{\rm max}$, but only by $\lsim 50\%$ \citep[][]{Laursen13}. A useful way to gain intuition on EW is that EW$\sim$FWHM$\times$(relative peak flux density). That is, for typical observed (restframe) FWHM of Ly$\alpha$ lines of FWHM$\sim 1-2$ \AA, EW=$240$ \AA\hs corresponds to having a relative flux density in the peak of the line that is $\sim 100$ times that in the continuum.

\subsection{Collisionally-excited (a.k.a `Cooling') Radiation}
\label{sec:co}
\begin{figure}[t]
\vspace{-5.0mm}
\vbox{\centerline{\epsfig{file=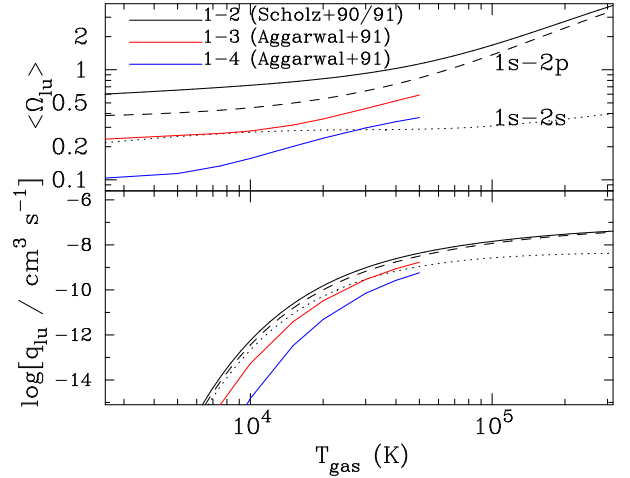,angle=0,width=10.0truecm}}}
\vspace{-5.0mm}
\caption[]{{\it Top panel}: velocity averaged collision strength $\langle \Omega_{lu}\rangle$ as a function of temperature $T$ for the Ly$\alpha$ (1s$\rightarrow$ 2p) transition ({\it black dashed line}). The {\it black dotted line} corresponds to the 1s$\rightarrow$ 2s transitions. {\it Solid lines} indicate transitions from the ground-state to excited states (summed over different orbital quantum numbers). {\it Bottom panel}: collision coupling $q_{\rm lu}$ (Eq~\ref{eq:qlu}) for the same transitions.}
\label{fig:omega}
\end{figure}  

Ly$\alpha$ photons can also be produced following collisional-excitation of the $2p$ transition when a hydrogen atoms deflects the trajectory of an electron that is passing by. The Ly$\alpha$ emissivity following collisional-excitation is given by
\begin{equation}
j_{{\rm coll}}(n,T,\nu)=\phi(\nu)n_{\rm e} n_{\rm HI}q_{\rm 1s2p} E_{\alpha}/(\sqrt{\pi}\Delta \nu_{\rm D}),
\end{equation} where $n_{\rm HI}$ denotes the number density of hydrogen atoms and 
\begin{equation}
q_{1s2p}(T)=8.63\times
10^{-6}T^{-1/2}\langle\Omega_{1s2p}\rangle \exp\Big{(}
-\frac{E_{\alpha}}{k_BT}\Big{)}\hs {\rm cm}^{3\hs}{\rm s}^{-1}
\label{eq:qlu}
\end{equation} where $\langle\Omega_{1s2p}\rangle(T)$ denotes the velocity averaged collision strength, which depends weakly on temperature. The {\it top panel} of Figure~\ref{fig:omega} shows the temperature dependence of $\langle \Omega_{lu}\rangle$ for the $1s\rightarrow 2s$  ({\it dotted line}), $1s\rightarrow 2p$ ({\it dashed line}), and for their sum $1s \rightarrow 2$ ({\it black solid line}) as given by \citet{Scholz90,Scholz91}. Also shown are velocity averaged collision strengths  for the $1s\rightarrow 3$ ({\it red solid line}, obtained by summing over all transitions $3s,3p$ and $3d$), and $1s\rightarrow 4$ ({\it blue solid line}, obtained by summing over all transitions $4s,4p,4d$ and $4f$) as given by \citet{Aggarwal91}. The {\it bottom panel} shows the collision coupling parameter $q_{\rm lu}$ for the same transitions. This plot shows that collisional coupling to the $n=2$ level increases by $\sim 3$ orders of magnitude magnitude when $T=10^4 K$ $\rightarrow$ $T=2\times 10^4$ K. The actual production rate of Ly$\alpha$ photons can be even more sensitive to $T$, as both $n_{\rm e}$ sharply increases with $T$ and $n_{\rm HI}$ sharply decreases with $T$ within the same temperature range (under the assumption that collisional ionisation balances recombination, which is relevant in e.g self-shielded gas, see e.g. Fig~1 in Thoul \& Weinberg 1995).

This process converts thermal energy of the gas into radiation, and therefore cools the gas. Ly$\alpha$ cooling radiation has been predicted to give rise to spatially extended Ly$\alpha$ radiation \citep[][]{Haiman00,Fardal01}, and provides a possible explanation for Ly$\alpha$ `blobs' \citep[][]{DL09,Goerdt10,FG10,BR12}. In these models, the Ly$\alpha$ cooling balances `gravitational heating' in which gravitational binding energy is converted into thermal energy in the gas. 

Precisely how gravitational heating works is poorly understood. \citet{Haiman00} propose that the gas releases its binding energy in a series of `weak' shocks as the gas navigates down the gravitational potential well. These weak shocks convert binding energy into thermal energy over a spatially extended region, which is then reradiated primarily as Ly$\alpha$. It is possible that a significant fraction of the gravitational binding energy is released very close to the galaxy \citep[e.g. when gas free-falls down into the gravitational potential well, until it is shock heated when it `hits' the galaxy][]{BD03}. It has been argued that some compact Ly$\alpha$ emitting sources may be powered by cooling radiation \citep[as in][]{BD03,Dcool,Dayal10}. Recent hydrodynamical simulations of galaxies indicate that the fraction of Ly$\alpha$ flux coming from galaxies in the form of cooling radiation increases with redshift, and may be as high as $\sim 50\%$ at $z\sim 6$ \citep[][]{Dayal10,Yajima12}. However, one should take these numbers with caution, because the predicted Ly$\alpha$ cooling luminosity depends sensitively on the gas temperature of the `cold' gas  (i.e. around T$\sim 10^4$ K, as illustrated by the discussion above). It is very difficult to reliably predict the temperature of this gas, because the gas' short cooling time drives the gas temperature to a value where its total cooling rate balances its heating rate. Because of this thermal equilibrium, we must accurately know and compute all the heating rates in the ISM \citep[][]{FG10,C12,BR12} to make a robust prediction for the Ly$\alpha$ cooling rate. These heating rates include for example photoionization heating, which requires coupled radiation-hydrodynamical simulations \citep[as][]{BR12}, or shock heating by supernova ejecta \citep[e.g.][]{ShullMcKee}. 

It may be possible to observationally constrain the contribution of cooling radiation to the Ly$\alpha$ luminosity of a source, through measurements of the Ly$\alpha$ equivalent width: the larger the contribution from cooling radiation, the larger the EW. Ly$\alpha$ emission powered by regular star-formation can have EW$_{\rm max}\sim 300-400$ \AA (see discussion above). Naturally, observations of Ly$\alpha$ emitting galaxies whose EW significantly exceeds EW$_{\rm max}$ \citep[as in e.g.][]{K12}, may provide hints that we are detecting a significant contribution from cooling.  However, the same signature can be attributed population III stars \citep[e.g.][]{Raiter}, and/or galaxies forming stars with a top-heavy initial mass function \citep[IMF, e.g.][]{MR}, or stochastic sampling of the IMF \citep{Jaime}. In theory one can distinguish cooling radiation from these other processes via the Balmer lines, because Ly$\alpha$ cooling radiation is accompanied by an H$\alpha$ luminosity that is $\frac{j_{{\rm coll,Ly}\alpha}}{j_{{\rm coll,H}\alpha}}=\frac{E_{\alpha}}{E_{{\rm H}\alpha}}\frac{\langle \Omega_{\rm 1s2p}\rangle}{\langle \Omega_{\rm 13}\rangle}$$\exp\big{(}\frac{E_{{\rm H}\alpha}}{k_{\rm B} T}\big{)}$ $\sim 100$ times weaker, which is much weaker than expected for case-B recombination \citep[where the H$\alpha$ flux is $\sim 8$ times weaker, e.g.][]{DL09}. Measuring the flux in the H$\alpha$ line at these levels requires an IR spectrograph with a sensitivity comparable to that of JWST\footnote{For example, for a Ly$\alpha$ source with an intrinsic luminosity of $L_{\alpha}=10^{43}$ erg/s at $z=5$, the corresponding H$\alpha$ flux is $F_{{\rm H}\alpha}=4.4\times 10^{-18}f_{\rm esc}^{H\alpha}\Big{(}\frac{F_{{\rm Ly}\alpha}/F_{{\rm H}\alpha}}{8.7}\Big{)}^{-1}$ erg/s/cm$^2$, which is too faint to be detected with existing IR spectrographs (here $f_{\rm esc}^{H\alpha}$ denotes the escape fraction of H$\alpha$ photons). However, these flux levels can be reached at $\sim 40\sigma \Big{(}\frac{F_{{\rm Ly}\alpha}/F_{{\rm H}\alpha}}{8.7}\Big{)}^{-1}$ in $10^4$ sec with NIRSPEC on JWST, provided that the flux is in an unresolved point source (see http://www.stsci.edu/jwst/science/sensitivity/).}.
\begin{figure*}
\vspace{-10mm}
\vbox{\centerline{\epsfig{file=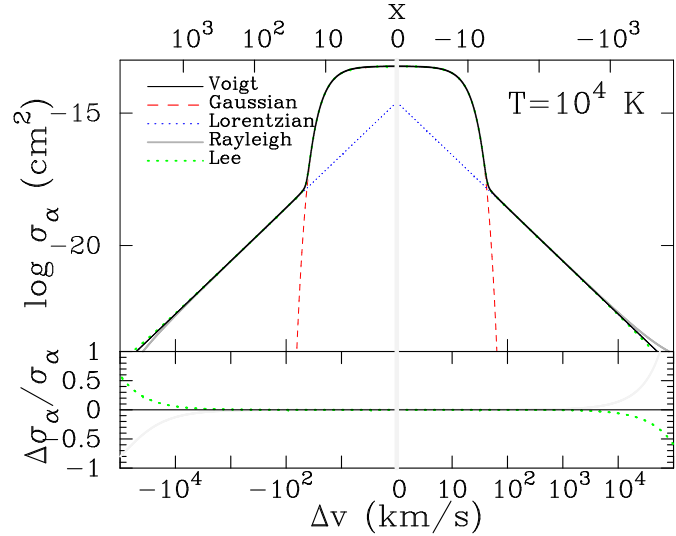,angle=0,width=10.9truecm}}}
%\centerline{\includegraphics[width=11.5cm]{cross.pdf}}
\vspace{-5mm}
\caption[]{The {\it black solid line} in {\it top panel} shows the Ly$\alpha$ absorption cross section, $\sigma_{\alpha}(x)$, at a gas temperature of $T=10^4$ K as given by the Voigt function (Eq~\ref{eq:voigt}). This Figure shows that the absorption cross section is described accurately by a Gaussian profile ({\it red dashed line}) in the `core' at $|x|<x_{\rm crit}\sim 3.2$ (or $|\Delta v| < 40$ km s$^{-1}$), and by a Lorentzian profile in the `wing' of the line ({\it blue dotted line}).  The Voigt profile is only an approximate description of the real absorption profile. Another approximation includes the `Rayleigh' approximation ({\it grey solid line}, see text). The {\it green dotted line} shows the absorption profile resulting from a full quantum mechanical calculation (Lee 2013). The different cross sections are compared in the {\it lower panel}, which highlights that the main differences arise only far in the wings of the line.}
\label{fig:sigma}
\end{figure*} 
\subsection{Boosting Recombination Radiation}
Equation~\ref{eq:0.68} was derived assuming case-B recombination. However, at $Z \lsim 0.03 Z_{\odot}$ significant departures from case-B are expected. These departures increase the Ly$\alpha$ luminosity relative to case-B \citep[e.g.][]{Raiter}. This increase of the Ly$\alpha$ luminosity towards lower metallicities is due to two effects: ({\it i}) the increased temperature of the HII region as a result of a suppressed radiative cooling efficiency of metal-poor gas. The enhanced temperature in turn increases the importance of collisional processes. For example, collisional-excitation increases the population of H-atoms in the $n=2$ state, which can be photoionized by lower energy photons. Moreover, collisional processes can mix the populations of atoms in their $2s$ and $2p$ states; ({\it ii}) harder ionizing spectra emitted by metal poor(er) stars. Higher energy photons can in principle ionize more than 1 H-atom, which can boost the overall Ly$\alpha$ production per ionizing photon. \cite{Raiter} provide a simple analytic formula which capture all these effects:
\begin{equation}
\dot{N}^{\rm rec}_{{\rm Ly}\alpha}= f_{\rm coll} P(1-\fesc)\dot{N}_{\rm ion},
\label{eq:raiter}
\end{equation}

where $P\equiv \langle E_{\gamma,{\rm ion}} \rangle/13.6\hs{\rm eV}$, in which $\langle E_{\gamma,{\rm ion}} \rangle$ denotes the mean energy of ionising photons\footnote{That is, $ \langle E_{\gamma,{\rm ion}} \rangle\equiv h\frac{\int_{13.6\hs{\rm eV}}^{\infty}d\nu f(\nu)}{\int_{13.6\hs{\rm eV}}^{\infty}d\nu f(\nu)/\nu}$, where $f(\nu)$ denotes the flux density.}. Furthermore, $f_{\rm coll}\equiv \frac{1+an_{\rm HI}}{b+cn_{\rm HI}}$, in which $a=1.62\times 10^{-3}$, $b=1.56$, $c=1.78\times 10^{-3}$, and $n_{\rm HI}$ denotes the number of density of hydrogen nuclei. Eq~\ref{eq:raiter} resembles the `standard' equation, but replaces the factor 0.68 with $Pf_{\rm coll}$, which can exceed unity. Eq~\ref{eq:raiter} implies that for a fixed IMF, the Ly$\alpha$ luminosity may be boosted by a factor of a few. Incredibly, for certain IMFs the Ly$\alpha$ line may contain $40\%$ of the total bolometric luminosity of a galaxy, which corresponds to a rest frame EW$\sim 4000$ \AA. 

We point out that the collisional processes discussed here are distinct from the collisional-excitation process discussed above (in \S~\ref{sec:co}), as they do not {\it directly} produce Ly$\alpha$ photons. Instead, they boost the number of Ly$\alpha$ photons that we can produce per ionising photon.

\section{Ly$\alpha$ Radiative Transfer Basics}
\label{sec:transfer}

Ly$\alpha$ radiative transfer consists of absorption followed by (practically) instant reemission, and hence closely resembles pure scattering. Here, we review the basic radiative transfer that is required to understand why \& how Ly$\alpha$ emitting galaxies probe the EoR.

It is common to express the frequency of a photon $\nu$ in terms of the dimensionless variable $x\equiv (\nu-\nu_{\alpha})/\Delta \nu_{\rm D}$. Here, $\nu_{\alpha}=2.46 \times 10^{15}$ Hz denotes the frequency corresponding the Ly$\alpha$ resonance, and $\Delta \nu_{\rm D} \equiv \nu_{\alpha}\sqrt{2kT/m_p  c^2}\equiv \nu_{\alpha}v_{\rm th}/c$. Here, $T$ denotes the temperature of the gas that is scattering the Ly$\alpha$ radiation, and $v_{\rm th}$ denotes the thermal speed.

\subsection{The Cross Section}
\label{sec:cross}

The frequency dependence of the Ly$\alpha$ absorption cross-section, $\sigma_{\alpha}(x)$, is described well by a Voigt function. That is
\begin{align}
\label{eq:voigt}
\sigma_{\alpha}(x)=\sigma_0\times \frac{\av}{\pi} \int_{-\infty}^{+\infty}dy \frac{\exp(-y^2)}{(x-y)^2+\av^2}\equiv \sigma_0\times \phi(x).\\ \nonumber
\sigma_0= \frac{f_{\alpha}}{\sqrt{\pi}\Delta \nu_{\rm D}}\frac{\pi e^2}{m_e c}=5.88 \times 10^{-14}(T/10^4\hs {\rm K})^{-1/2}\hs {\rm cm}^{2}
\end{align} where $f_{\alpha}=0.416$ denotes the Ly$\alpha$ oscillator strength, and $\av=A_{\alpha}/[4\pi \Delta \nu_{\rm D}]=4.7 \times 10^{-4}(T/10^{4}\hs{\rm K})^{-1/2}$ denotes the Voigt parameter, and $\sigma_0$ denotes the cross section at line center. We introduced the Voigt function\footnote{We adopt the normalization $\phi(0)=1$, which translates to $\int \phi(x)dx=\sqrt{\pi}$.} $\phi(x)$, which is plotted as the {\it black solid line} in the {\it upper panel} of Figure~\ref{fig:sigma}. This Figure also shows that the Voigt function $\phi(x)$ is approximated accurately as
\begin{eqnarray}
\phi(x)\approx \left\{ \begin{array}{ll}
         \ e^{-x^2}& \mbox{`core', i.e.}\hs |x|<x_{\rm crit};\\
         \ \frac{\av}{\sqrt{\pi}x^2}& \mbox{`wing', i.e.}\hs |x|>x_{\rm crit},\end{array} 
\right. 
\label{eq:sigma}
\end{eqnarray} where the transition from the Gaussian core ({\it red dashed line}) to the Lorentzian wing ({\it blue dotted line})\footnote{The Ly$\alpha$ absorption line profile of an individual HI atom is given by a Lorentzian function, $\phi_{\rm L}(x)\propto (x^2+\av^2)^{-1}$. This Lorentzian profile is also plotted. The figure clearly shows that far from line center, we effectively recover the single atom or Lorentzian line profile.} occurs at $x_{\rm crit} \sim 3.2$ at the gas temperature of $T=10^4$ K that we adopted. An even more accurate fit - which works well even in the regime where we transition from core to wing - is given in \citet{Ta06}.

It is important to point out that the Voigt function itself (as given by Eq~\ref{eq:voigt}) only represents an approximation to the real frequency dependence of the absorption cross section. The Voigt profile is derived through a convolution of a Gaussian profile (describing the thermal velocity-distribution of HI atoms) with the Lorentzian profile (see above). A common modification of the Lorentzian is given in e.g. \citet{Peebles93,Peebles69}, where the absorption cross section for a single atom includes an additional $(\nu/\nu_{\alpha})^4$-dependence, as is appropriate for Rayleigh scattering\footnote{This additional term arises naturally in a classical calculation in which radiation of frequency $\nu$ scatters off an electron that orbits the proton at a natural frequency $\nu_0$.}. In this approximation we have \citep[also see][]{Schroeder}
\begin{align}
\label{eq:ray}
\sigma_{\alpha,{\rm Ray}}(x)=\sigma_{\alpha}(x)\Big{(}\frac{\nu}{\nu_{\alpha}} \Big{)}^4=\sigma_{\alpha}(x)(1+xv_{\rm th}/c)^4,
\end{align} which gives rise to slightly asymmetric line profiles (as shown by the {\it grey solid line} in Fig~\ref{fig:sigma}). However, even this still represents an approximation \citep[as pointed out in][]{Peebles69}. A complete quantum-mechanical derivation of the frequency-dependence of the Ly$\alpha$ absorption cross section has been presented only recently by \citet{Lee13}, which can be captured by the following correction to the Voigt profile:
\begin{align}
\sigma_{\alpha,{\rm Lee}}(x)=\sigma_{\alpha}(x)(1-1.792xv_{\rm th}/c).
\end{align} This cross section is shown as the {\it green dotted line}. In contrast to the Rayleigh-approximation given above, the red wing is strengthened relative to the pure Lorentzian proÞle, which \citet{Lee13} credits to positive interference of scattering from all other levels. The {\it lower panel} shows the fractional difference of the three cross sections $\sigma_{\alpha,{\rm Ray}}(x)$, $\sigma_{\alpha,{\rm Lee}}(x)$, and $\sigma_{\alpha}(x)$.\\
 
Although the Voigt function does not capture the full frequency dependence of Ly$\alpha$ absorption cross section far in the wings on the line, it clearly provides an accurate description of the cross section near the core. The fast reduction in the cross section outside of the core (here at $|\Delta v| > 10$ km s$^{-1}$) enables Ly$\alpha$ photon to escape more easily from galaxies, which - with HI column densities in excess of $N_{\rm HI}=10^{20}$ cm$^{-2}$ - are extremely opaque to Ly$\alpha$ photons. This process is discussed in more detail in the next sections.

\subsection{Frequency Redistribution $R(\nu',\nu)$: Resonance vs Wing Scattering}
\label{sec:redist}
  
Absorption of an atom is followed by re-emission on a time scale $A^{-1}_{2p1s}\sim 10^{-9}$ s. At interstellar and intergalactic densities, atoms are very likely not `perturbed' in such short times scales\footnote{The rate at which atoms in the $2p$ state interact with protons is given by $C_{\rm 2p2s}n_{\rm p}\sim 1.6 \times 10^{-4} n_{\rm p}$ s$^{-1}$. For electrons the corresponding rate is reduced by a factor of $\sim 10$. The interaction with other neutral atoms is negligible.}, and the atom {\it re-emits a photon with an energy that equals that of the absorbed photon} when measured in the atom's frame. Because of the atom's thermal motion however, in the lab frame the photon's energy will be Doppler boosted. The photon's frequency before and after scattering are therefore not identical but correlated, and the scattering is referred to as `partially coherent' (completely coherent scattering would refer to the case where the photons frequency before and after scattering are identical). 

\begin{figure}[t]
\vbox{\centerline{\epsfig{file=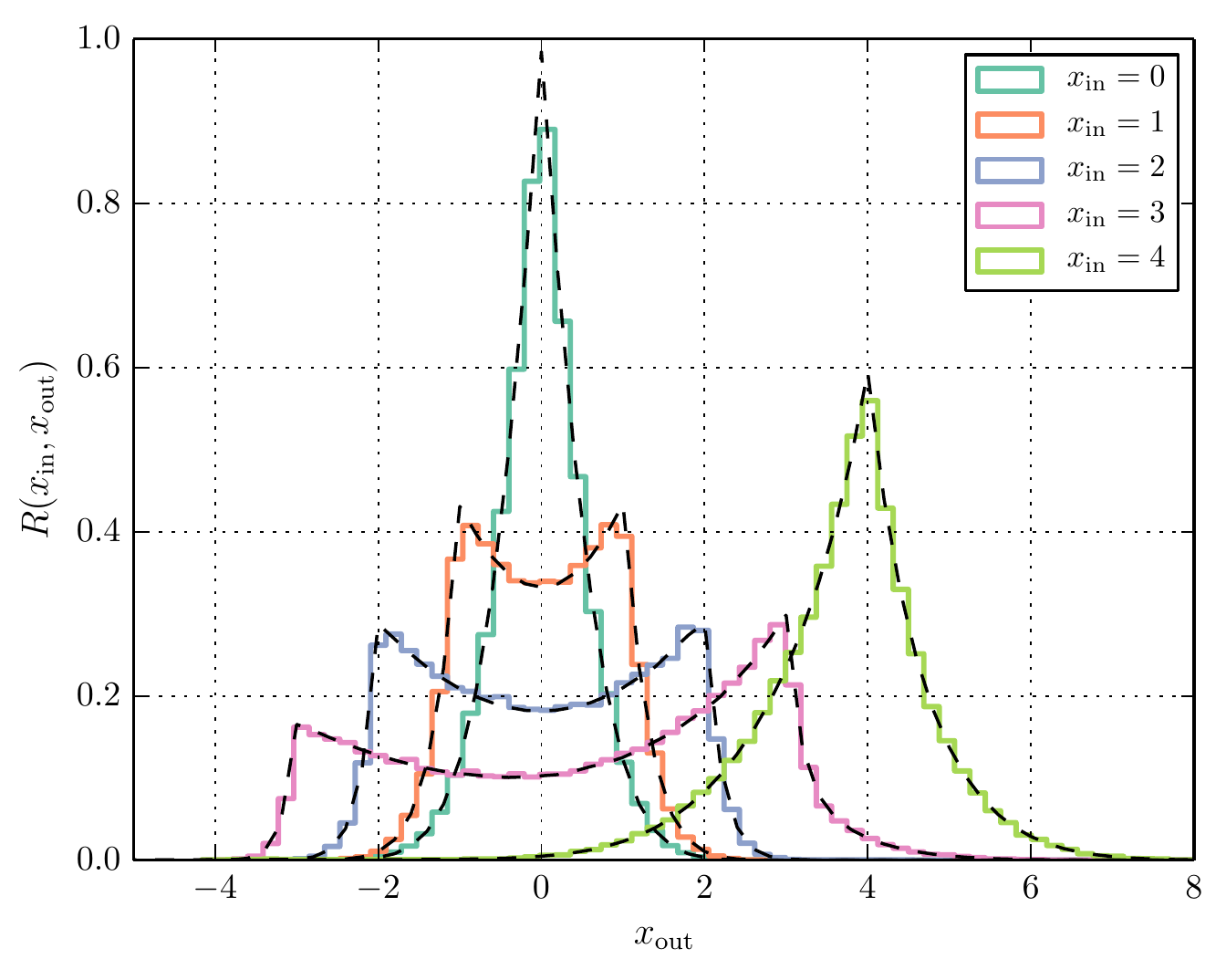,angle=0,width=8.0truecm}}}
\vspace{5.0mm}
\caption[]{This Figure ({\it Credit: Figure kindly provided by Max Gronke}) shows examples of redistribution functions - the PDF of the frequency of the photon after scattering ($\xo$, here labelled as x'), given its frequency before scattering ($ \xin$, here labelled as $x$) - for partially coherent scattering. We show cases for $\xin=0,1,2,3,4,5$. The plot shows that photons in the wing (e.g. at $\xin=5$) are unlikely to be scattered back into the core in a single scattering event.}
\label{fig:redist}
\end{figure}  

We can derive a probability distribution function (PDF) for the photons frequency after scattering, $\xo$, given its frequency before scattering, $\xin$. Expressions for these `redistribution functions'\footnote{It is worth pointing out that these redistribution functions are {\it averaged} over the direction in which the outgoing photon is emitted, i.e.
\begin{equation} 
R(\xo,\xin)=\frac{1}{4\pi}\int_{-1}^{1}d\mu \hs P(\mu)R(\xo,\xin,\mu),
\end{equation} where $P(\mu)$ denotes the `phase function', and $P(\mu)d\mu/[4\pi]$ describes the probability that $\mu = {\bf k}_{\rm out} \cdot {\bf k}_{\rm in}$ lies in the range $\mu \pm d\mu/2$. We stress that the redistribution functions depend strongly on outgoing direction. Expressions for $R(\xo,\xin,\mu)$ can be found in Dijkstra \& Kramer (2012).} can be found in e.g. Lee (1974, also see Unno 1952, Hummer 1962). Redistribution functions that describe partially coherent scattering have been referred to as `type-II' redistribution functions (where type-I would refer to completely incoherent scattering).

Figure~\ref{fig:redist} shows examples of type-II redistribution functions, $R(\xo,\xin)$, as a function of $\xo$ for $\xin=0,1,...$ (this Figure was kindly provided by Max Gronke). This Figure shows that ({\it i}) $R(\xo,\xin)$ varies rapidly with $\xin$, and ({\it ii}) for $|\xin| \gg 3$ the probability of being scattered back to $\xo=0$ becomes vanishingly small. Before we discuss why this latter property of the redistribution function has important implications for the scattering process, we first explain that its origin is related to `resonant' vs `wing' scattering.

\begin{figure}[t]
\vspace{-10mm}
\vbox{\centerline{\epsfig{file=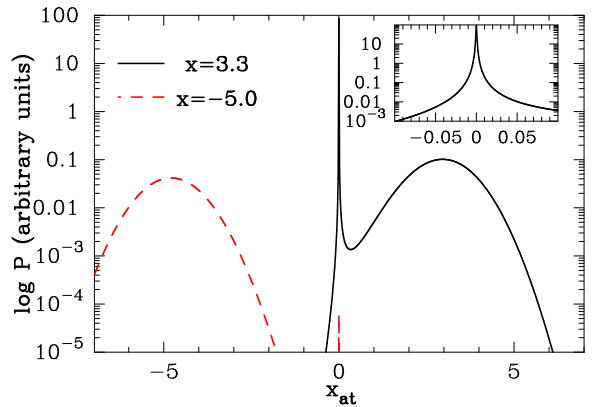,angle=0,width=9.0truecm}}}
\vspace{-5mm}
\caption[]{The probability that a photon of frequency $\xin$ is scattered by an atom such that it appears at a frequency $x_{\rm at}$ in the frame of the atom (here $\xin$ is labelled as $x$, {\it Credit: from Figure~{\bf A.2} of Dijkstra \& Loeb 2008, `The polarization of scattered Ly$\alpha$ radiation around high-redshift galaxies', MNRAS, 386, 492D}). The
{\it solid}/{\it dashed line} corresponds to $\xin=3.3$/$\xin=-5.0$ For $\xin=3.3$, photons are either scattered by atoms to which they appear exactly at resonance (see the {\it inset}, which shows the region around $x_{\rm at}$ in more detail) - hence 'resonant' scattering - or to which they appear $\sim 3$ Doppler widths away. For $\xin=-5$ the majority of photons scatter off atoms to which they appear in the wing.}
\label{fig:pvel}
\end{figure}

Figure~\ref{fig:pvel} shows the PDF of the frequency of a photon, in the atoms frame ($x_{\rm at}$), for two incoming frequency $\xin=3.3$ ({\it black solid line}, here labelled as $x$) and $\xin=-5.0$ ({\it red dashed line}). The {\it black solid line} peaks at $x_{\rm at}=0$. That is, the photon with $\xin=3.3$ is most likely scattered by an atom to which the photon appears exactly at line centre. That is, the scattering atom must have velocity component parallel to the incoming photon that is $\sim 3.3$ times $v_{\rm th}$. This requires the atom to be on the Maxwellian tail of the velocity distribution. Despite the smaller number of atoms that can meet this requirement, there are still enough to dominate the scattering process. However, when $\xin=-5.0$ the same process would require atoms that lie even further on the Maxwellian tail. These atoms are too rare to contribute to scattering. Instead, the photon at $\xin=-5.0$ is scattered by the more numerous atoms with speeds close to $v_{\rm th}$. In the frame of these atoms, the photon will appear centered on $x_{\rm at}=\xin=-5.0$ (as shown by the {\it red dashed line}).

As shown above, a fraction of photons at $\xin=3.3$ scatter off atoms to which they appear very close to line centre. Thus, a fraction of these photons scatter `resonantly'. In contrast, this fraction is vanishingly small for the photons at $\xin=-5.0$ (or more generally, for all photons with $|\xin| \gg 3$). These photons do not scatter in the wing of the line. This is more than just semantics, the phase-function and polarisation properties of scattered radiation depends sensitively on whether the photon scattered resonantly or not \citep[][]{Stenflo80,RL99,Dpol}.\\

%-----
There are two useful \& important expectation values of the redistribution functions for photons at $|\xin| \gg 3$ \citep[e.g.][]{Osterbrock62,FP06}:
\begin{eqnarray}
E(\sqrt{\Delta x^2}|\xin)=1 & \\ \nonumber
E(\Delta x |\xin)=-\frac{1}{\xin} & |\xin|\gg 1,
\label{eq:redist}
\end{eqnarray} where $\Delta x \equiv \xo -\xin$, and the expectation values are calculated as $E(X|\xin) \equiv$ $\int_{-\infty}^{\infty}d\xo \hs X R(\xo,\xin)$.
%-----

The first equality states that the r.m.s. frequency change of the photon before and after scattering equals 1 Doppler width. This is an important result: a photon that is absorbed far in the wing of the line, will remain far in the wing after scattering, which facilitates the escape of photons (see below).  The second equality states that for photons that are absorbed in the wing of the line, there is a slight tendency to be scattered back to the core, e.g. a photon that was at $\xin=10$, will have an outgoing frequency around $\langle \xo \rangle =9.9$. The second equality also implies that photons at $|x|\gg 3$ typically scatter $N_{\rm scat }\sim x^2$ times before they return to the core (in a static medium). These photons can travel a distance $\sqrt{N_{\rm scat}}\lambda_{\rm mfp}(x)\sim |x|\lambda_{\rm mfp}(x)\propto |x|/\phi(x)$ from where they were emitted. This should be compared with the distance $\lambda_{\rm mfp}(x) \propto 1/\phi(x)$ that can be travelled by photons at $|x| \lsim 3$. The path of photons in real space as they scatter in the wing of the line (i.e. at $|x| \gg 3$) back towards the core is referred to as an `excursion'. The optical depth of a static uniform medium beyond which photons preferably escape in `excursions' marks the transition from `optical thick' to `extremely optical thick'.
Finally, the second equality also allows us to estimate the spectrum of Ly$\alpha$ photons that escape from an extremely opaque, static medium as is discussed in \S~\ref{sec:static} below.

\subsection{Ly$\alpha$ Scattering in Static Media}
\label{sec:static}
\begin{figure}[t]
\vspace{-15.0mm}
\vbox{\centerline{\epsfig{file=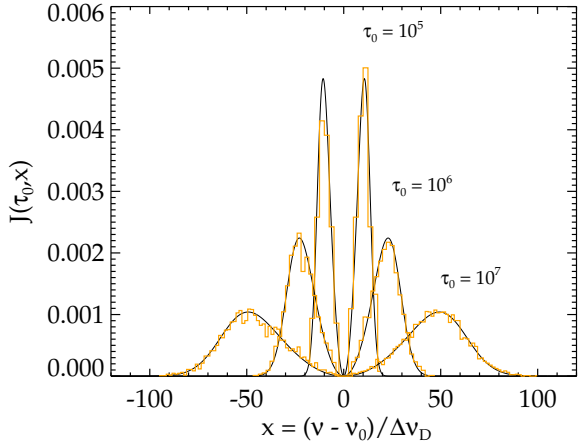,angle=0,width=11.0truecm}}}
\vspace{-10.0mm}
\caption[]{Ly$\alpha$ spectra emerging from a uniform spherical, static gas cloud surrounding a central Ly$\alpha$ source which emits photons at line centre $x=0$. The total line-center optical depth, $\tau_0$ increases from $\tau_0=10^5$ ({\it narrow histogram)} to $\tau_0=10^7$ ({\it broad histogram}). The {\it solid lines} represent analytic solutions ({\it Credit: from Figure~ {\bf A2} of Orsi et al. 2012, `Can galactic outflows explain the properties of Ly$\alpha$ emitters?', MNRAS, 425, 87O)}.}
\label{fig:static}
\end{figure} 
Consider of source of Ly$\alpha$ photons in the center of a static, homogeneous sphere, whose line-center optical depth from the center to the edge equals $\tau_0$, where $\tau_0$ is extremely large (say $\tau_0=10^7$). We further assume that the central source emits all Ly$\alpha$ photons at line center. The photons initially resonantly scatter in the core of the line profile, with a mean free path that is negligible small compared to the size of the sphere. Because the photons change their frequency during each scattering event, the photons 'diffuse' in frequency space as well. We expect on rare occasions the Ly$\alpha$ photons to be scattered into the wing of the line. The mean free path of a wing photon at frequency $x$ equals $1/\phi(x)$ in units of line-center optical depth. Photons that are in the wing of the line scatter $N_{\rm scat} \sim x^2$ times before returning to the core, but will have diffused a distance $\sim \sqrt{N_{\rm scat}}/\phi(x)$ from the center of the sphere. If we set this displacement equal to the size of the sphere, i.e. $N_{\rm scat}/\phi(x)=\tau_0$, and solve for $x$ using that $\phi(x)=\av/[\sqrt{\pi}x^2]$, we find $x_{\rm p}= \pm [\av\tau_0/\sqrt{\pi}]^{1/3}$ \citep{Adams72,Harrington73,Neufeld90}. Photons that are scattered to frequencies\footnote{Apart from a small recoil effect that can safely ignored (Adams 1971), photons are equally likely to scatter to the red and blue sides of the resonance.} $|x| < |x_{\rm p}|$ will return to line center before they escape from the sphere (where they have negligible chance to escape). Photons that are scattered to frequencies $|x| > |x_{\rm p}|$ can escape more easily, but there are fewer of these photons because: ({\it i}) it is increasingly unlikely that a single scattering event displaces the photon to a larger $|x|$, and ({\it ii}) photons that wish to reach $|x|\gg |x_p|$ through frequency diffusion via a series of scattering events are likely to escape from the sphere before they reach this frequency. 

We therefore expect the spectrum of Ly$\alpha$ photons emerging from the center of an extremely opaque object to have two peaks at $x_{\rm p} =\pm k[\av\tau_0/\sqrt{\pi}]^{1/3}$, where $k$ is a constant of order unity which depends on geometry (i.e. $k=1.1$ for a slab [Harrington 1973, Neufeld 1990], and $k=0.92$ for a sphere [Dijkstra et al. 2006]). This derivation required that photons escape in a single excursion. That is, photons must have been scattered to a frequency $|x|\gg 3$ (see \S~\ref{sec:redist}). If for simplicity we assume that $x_{\rm p}\sim x$ then escape in a single excursion - and hence the transition to extremely opaque occurs - when $x_{\rm p} \gg 3$ or when $\av\tau_0 = \sqrt{\pi}(x_{\rm p}/k)^3\gsim 1600 (x_{\rm p}/10)^3$. Indeed, analytic solutions of the full spectrum emerging from static optically thick clouds appear in good agreement with full Monte-Carlo calculations (see \S~\ref{sec:mc}) when $\av\tau_0 \gsim 1000$ \citep[e.g.][]{Neufeld90,D06}. 

These points are illustrated in Figure~\ref{fig:static} where we show spectra of Ly$\alpha$ photons emerging from static, uniform spheres of gas surrounding a central Ly$\alpha$ source (taken from Orsi et al. 2012, the assumed gas temperature is $T=10$ K). This Figure contains three spectra corresponding to different $\tau_0$. {\it Solid lines}/{\it histograms} represent spectra obtained from analytic calculations/ Monte-Carlo simulations. This Figure shows the spectra contain two peaks, located at $x_{\rm p}$ given above. The Monte-Carlo simulations and the analytic calculations agree well. For $T=10$ K, we have $\av=1.5\times 10^{-2}$ and $\av\tau_0=1.5\times 10^3$, and we expect photons to escape in a single excursion, which is captured by the analytic calculations. \\

We can also express the location of the two spectral peaks in terms of a velocity off-set and an HI column density as

\begin{equation}
\Delta v_{\rm p}\approx 160\Big{(}\frac{N_{\rm HI}}{10^{20}\hs {\rm cm^{-2}}} \Big{)}^{1/3}\Big{(} \frac{T}{10^4\hs{\rm K}}\Big{)}^{-1/6}\hs{\rm km}\hs{\rm s}^{-1}.
\end{equation} That is, the full-width at half maximum of the Ly$\alpha$ line can exceed $2\Delta v_{\rm p} \sim 300$ km s$^{-1}$ for a static medium.

\subsection{Ly$\alpha$ Scattering in an Expanding/Contracting Medium}
\label{sec:exp}

\begin{figure*}[t]
\vspace{-125.0mm}
\vbox{\centerline{\epsfig{file=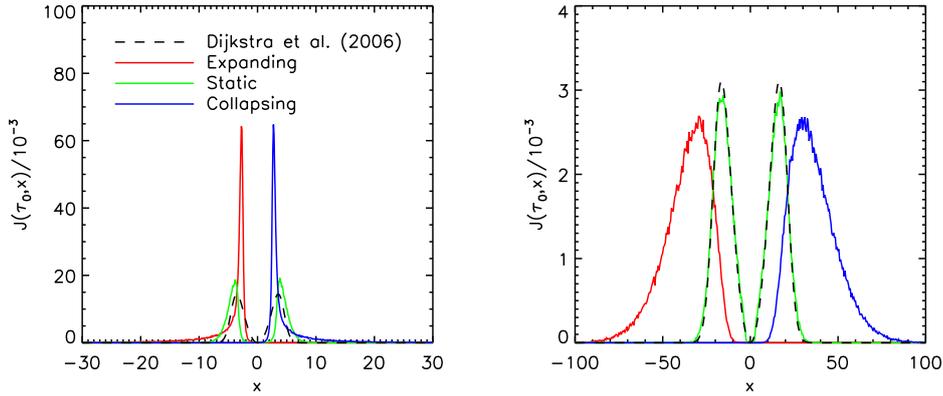,angle=0,width=14.0truecm}}}
%\vbox{\centerline{\epsfig{file=point_thick.eps,angle=0,width=8.0truecm}}}
\vspace{-3.0mm}
\caption[]{This Figure illustrates the impact of bulk motion of optically thick gas to the emerging Ly$\alpha$ spectrum of Ly$\alpha$: The {\it green lines} show the spectrum emerging from a static sphere (as in Fig~\ref{fig:static}). In the {\it left/right panel} the HI column density from the centre to the edge of the sphere is $N_{\rm HI}=2\times 10^{18}$/$2\times 10^{20}$ cm$^{-2}$. The {\it red/blue lines} show the spectra emerging from an expanding/a contracting cloud. Expansion/contraction gives rise to an overall redshift/blueshift of the Ly$\alpha$ spectral line ({\it Credit: from Figure~7 of Laursen et al. 2009b \textcopyright AAS. Reproduced with permission}).}
\label{fig:inout}
\end{figure*} 
For an outflowing medium, the predicted spectral line shape also depends on the outflow velocity, $v_{\rm out}$. Qualitatively, photons are less likely to escape on the blue side (higher energy) than photons on the red side of the line resonance because they appear closer to resonance in the frame of the outflowing gas. Moreover, as the Ly$\alpha$ photons are diffusing outward through an expanding medium, they loose energy because the do 'work' on the outflowing gas \citep[][]{ZM02}. Both these effects combined enhance the red peak, and suppress the blue peak, as illustrated in Figure~\ref{fig:inout} (taken from Laursen et al. 2009b). In detail, how much the red peak is enhanced, and the blue peak is suppressed (and shifted in frequency directions) depends on the outflow velocity and the HI column density of gas. 

There exists one analytic solution to radiative transfer equation through an expanding medium: \citet{LR99} derived analytic expressions for the angle-averaged intensity $J(\nu,r)$ of Ly$\alpha$ radiation as a function of distance $r$ from a source embedded within a neutral intergalactic medium undergoing Hubble expansion\footnote{This calculation assumed that $\xo=\xin$ which corresponds to the special case of $T=0$.}.

Not unexpectedly, the same arguments outlined above can be applied to an inflowing medium: here we expect the blue peak to be enhanced and the red peak to be suppressed \citep[e.g.][]{D06,Barnes10}. It is therefore thought that the Ly$\alpha$ line shape carries information on the gas kinematics through which it is scattering. As we discuss in \S~\ref{sec:ISM}, the shape and shift of the Ly$\alpha$ spectral line profile has been used to infer properties of the medium through which they are scattering.

\subsection{Ly$\alpha$ Transfer through a Dusty (Multiphase) Medium}
\label{sec:dust}
\begin{figure}[t]
\begin{center}
\vspace{-90mm}
\includegraphics[width=11.5cm]{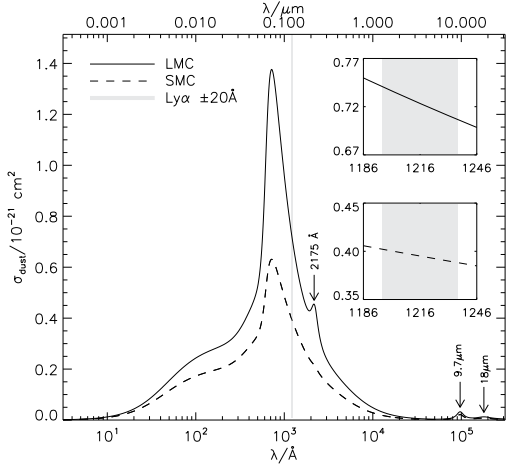}
\caption{This Figure shows grain averaged absorption cross section of dust grains {\it per hydrogen atom} for SMC/LMC type dust ({\it solid/dashed line}, see text). The {\it inset} shows the cross section in a narrower frequency range centered on Ly$\alpha$, where the frequency dependence depends linearly on $x$. This dependence is so weak that in practise it can be safely ignored ({\it Credit: from Figure~1 of Laursen et al. 2009a \textcopyright AAS. Reproduced with permission.}).}
\label{fig:dust}
\end{center}
\end{figure} 

Ly$\alpha$ photons can be absorbed by dust grains. A dust grain can re-emit the Ly$\alpha$ photon (and thus `scatter' it), or re-emit the absorbed energy of the Ly$\alpha$ photon as infrared radiation. The probability that the Ly$\alpha$ photon is scattered, and thus survives its encounter with the dust grain, is given by the `albedo' $A_{\rm dust} \equiv \frac{\sigma_{\rm scat}}{\sd}$, where $\sd$ denotes the total cross section for dust absorption, and $\sigma_{\rm scat}$ denotes the cross section for scattering. Both the albedo $A_{\rm dust}$ and absorption cross section $\sd$ depend on the dust properties. For example, \citet{Laursen09} shows that $\sd=4 \times 10^{-22}(Z_{\rm gas}/0.25Z_{\odot})$ cm$^{-2}$ for SMC type dust (dust encountered in the Small Magellanic Cloud),  and $\sd=7 \times 10^{-22}(Z_{\rm gas}/0.5Z_{\odot})$ cm$^{-2}$ for LMC (Large Magellanic Cloud) type dust. Here, $Z_{\rm gas}$ denotes the metallicity of the gas.  \citet{Laursen09} further show that the frequency dependence of the dust absorption cross section around the Ly$\alpha$ resonance can be safely ignored (see Figure~\ref{fig:dust}).

A key difference between a dusty and dust-free medium is that in the presence of dust, Ly$\alpha$ photons can be destroyed during the scattering process when $A_{\rm dust} < 1$.  Dust therefore causes the `escape fraction' ($\fesca$), which denotes the fraction Ly$\alpha$ photons that escape from the dusty medium, to fall below unity, i.e. $\fesca<1$. Thus, while scattering of Ly$\alpha$ photons off hydrogen atoms simply redistributes the photons in frequency space, dust reduces their overall number. Dust can similarly destroy continuum photons, but because Ly$\alpha$ photons scatter and diffuse spatially through the dusty medium, the impact of dust on Ly$\alpha$ and UV-continuum is generally different. In a uniform mixture of HI gas and dust, Ly$\alpha$ photons have to traverse a larger distance before escaping, which increases the probability to be destroyed by dust. In these cases we expect dust to reduce the EW of the Ly$\alpha$ line.\\

\begin{figure}[t]
\begin{center}
\includegraphics[width=7.5cm]{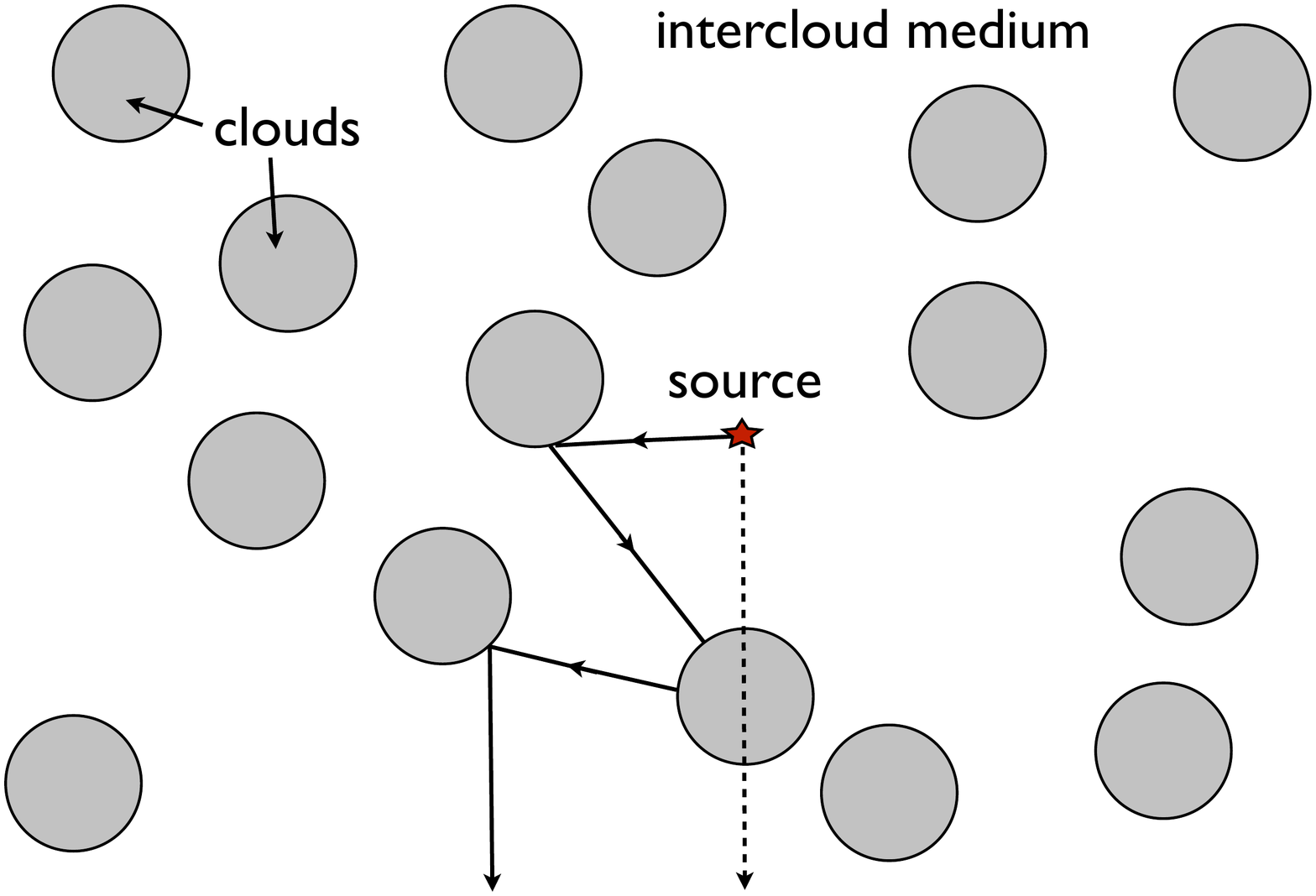}
\caption{Schematic illustration how a multiphase medium may favour the escape of Ly$\alpha$ line photons over UV-continuum photons: {\it Solid}/{\it dashed lines} show trajectories of Ly$\alpha$/UV-continuum photons through clumpy medium. If dust is confined to the cold clumps, then Ly$\alpha$ may more easily escape than the UV-continuum ({\it Credit: from Figure~1 of Neufeld 1991 \textcopyright AAS. Reproduced with permission.}).}
\label{fig:dust2}
\end{center}
\end{figure}
The presence of dust does not necessarily reduce the EW of the Ly$\alpha$ line. Dust can increase the EW of the Ly$\alpha$ line in a `clumpy' medium that consists of cold clumps containing neutral hydrogen gas and dust, embedded within a (hot) ionized, dust free medium \citep[][]{Neufeld91,Hansen06}.  In such a medium Ly$\alpha$ photons can propagate freely through the interclump medium, and scatter only off the surface of the neutral clumps, thus avoiding exposure to dust grains. In contrast, UV continuum photons will penetrate the dusty clumps unobscured by hydrogen and are exposed to the full dust opacity. This is illustrated visually in Figure~\ref{fig:dust2}. \citet{Laursen13} and \citet{Duval14} have recently shown however that EW boosting only occurs under physically unrealistic conditions in which the clumps are very dusty, have a large covering factor, have very low velocity dispersion and outflow/inflow velocities, and in which the density contrast between clumps and interclump medium is maximized. While a multiphase (or clumpy) medium definitely facilitates the escape of Ly$\alpha$ photons from dusty media, EW boosting therefore appears uncommon. We note the conclusions of \citet{Laursen13,Duval14} apply to the EW-boost averaged over {\it all} photons emerging from the dusty medium. \citet{GD14} have investigated that for a given model, there can be directional variations in the predicted EW, with large EW boosts occurring in a small fraction of sightlines in directions where the UV-continuum photon escape fraction was suppressed.

\subsection{Monte Carlo Radiative Transfer}
\label{sec:mc}

Analytic solutions to the radiative transfer equation (Eq~\ref{eq:RT}) only exist for a few idealised cases. A modern approach to solve this equation is via Monte-Carlo\footnote{It is worth pointing out that there exist numerous studies which explore alternatives to the Monte-Carlo approach, and focus on numerically solving approximations to the Ly$\alpha$ radiative transfer equation \citep[e.g.][]{Roy10,Yang11,HM12,Yang13}.}, in which scattering of individual photons is simulated until they escape \citep[e.g][]{LR99,Ahn01,ZM02,Cantalupo05,V06,Ta06,D06,Semelin07,Pierleoni09,K10,FG10,Barnes11,Zheng10,F11,Yajimacode,Orsi12,Behrens13}\footnote{The codes presented by \citet{Pierleoni09,Yajimacode} allow for simultaneous calculation of Ly$\alpha$ and ionising photons.}. Details on how the Monte-Carlo approach works can be found in many papers \citep[see e.g. the papers mentioned above, and Chapters~6-8 of][for an extensive description]{LaursenPhD}. To briefly summarise, for each photon in the Monte-Carlo simulations we first randomly draw a position, $r_{\rm i}$, at which the photon was emitted from the emissivity profile, a frequency $x_{\rm i}$ from the Voigt function $\phi(x)$, and a random propagation direction ${\bf k}_{\rm i}$. We then
\begin{enumerate}[leftmargin=*]
\item randomly draw the optical depth $\tau$ the photon propagates from the distribution $P(\tau)=\exp(-\tau)$.
\item convert $\tau$ to a physical distance $s$ by inverting the line integral $\tau =\int_0^s d\lambda n_{\rm HI} ({\bf r})\sigma_{\alpha}(x[{\bf r}])$, where ${\bf r} = {\bf r}_{\rm i}+\lambda{\bf k}$ and $x=x_{\rm i}-{\bf v}({\bf r})\cdot {\bf k}_{\rm i}/(v_{\rm th})$. Here, ${\bf v}({\bf r})$ denotes the 3D velocity vector of the gas at position ${\bf r}$.
\item randomly draw velocity components of the atom that is scattering the photon.
\item draw an outgoing direction of the photon after scattering, ${\bf k}_{\rm o}$, from the `phase-function', $P(\mu)$ where $\mu=\cos {\bf k}_{\rm i}\cdot{\bf k}_{\rm o}$. The functional form of the phase function depends on whether the photon is resonantly scattered or not (see \S~\ref{sec:redist}). It is worth noting that the process of generating the atom's velocity components and random new directions generates the proper frequency redistribution functions (as well as their angular dependence, see Dijkstra \& Kramer 2012). Unless the photon escapes, we replace the photons propagation direction \& frequency and go back to 1).
\end{enumerate}

Observables are then constructed by repeating this process many times and by recording the frequency (when predicting spectra), impact parameter and/or the location of the last scattering event (when predicting surface brightness profile\footnote{Predicting surface brightness profiles in gas distributions that are not spherically symmetric - as is generally the case - is more subtle, and requires the so-called `peeling' algorithm (also known as the `next-event estimator', see e.g. Laursen 2010 and references therein).}), and possibly the polarisation of each photon that escapes.

\section{Ly$\alpha$ Transfer in the Universe}
\label{sec:realRT}

In previous sections we summarised Ly$\alpha$ transfer in idealised optically thick media. In reality, the gas that scatters Ly$\alpha$ photons is more complex. 
Here, we provide a brief summary of our current understanding of Ly$\alpha$ transfer through the interstellar medium (\S~\ref{sec:ISM}), and also the post-reionization intergalactic medium (\S~\ref{sec:IGM}). For extensive discussion on these subjects, we refer to the reader to the reviews by Barnes et al. (2014), and by M. Hayes and S. Malhotra.

We purposefully make the distinction between intergalactic radiative transfer during and after reionization: in order to understand how reionization affects the visibility of Ly$\alpha$ from galaxies, it is important to understand how the intergalactic medium affects Ly$\alpha$ flux emitted by galaxies at lower redshift, and to understand how its impact evolves with redshift in the absence of diffuse neutral intergalactic patches that exist during reionization.

\subsection{Interstellar Radiative Transfer}
\label{sec:ISM}

To fully understand Ly$\alpha$ transfer at the interstellar level requires a proper understanding of the multiphase ISM, which lies at the heart of understanding star and galaxy formation. There exist several studies of Ly$\alpha$ transfer through simulated galaxies \citep{Ta06,Laursen07,Laursen09,Barnes11,Verhamme12,Yajima12}. It is important to keep in mind that modelling the neutral (outflowing) component of interstellar medium is an extremely challenging task, as it requires simultaneous resolving the interstellar medium and the impact of feedback by star-formation  on it (via supernova explosions, radiation pressure, cosmic ray pressure, ...). This requires (magneto)hydrodynamical simulations with sub-pc resolution \citep[e.g.][]{Cooper,Fujita09}. Instead of taking an `ab-initio' approach to understanding Ly$\alpha$ transfer, it is illuminating to use a `top-down' approach in which we try to constrain the broad impact of the ISM on the Ly$\alpha$ radiation field from observations. 

\begin{figure}[t]
\vspace{-110mm}
\hspace{15mm}
\vbox{\centerline{\epsfig{file=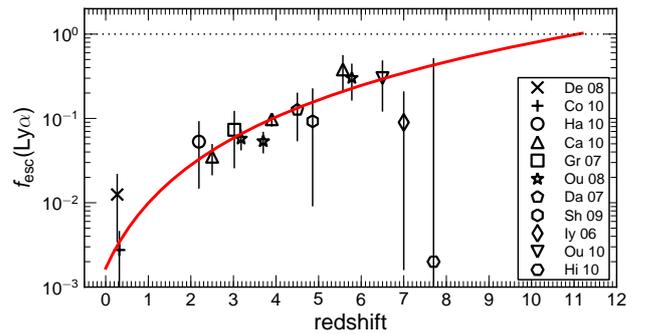,angle=0,width=12.0truecm}}}
\vspace{-5.0mm}
\caption[]{Observational constraints on the redshift-dependence of the volume averaged `effective' escape fraction, $\fesce$, which contains constraints on the true escape fraction $\fesca$ ({\it Credit: from Figure~1 of Hayes et al. 2011 \textcopyright AAS. Reproduced with permission}).}
\label{fig:fesc}
\end{figure}

Constraints on the escape fraction of Ly$\alpha$ photons, $\fesca$, have been derived by comparing the {\it intrinsic} Ly$\alpha$ luminosity, derived from (dust corrected) UV-derived and/or IR-derived star-formation rates of galaxies, to the observed Ly$\alpha$ luminosity. These analyses have revealed that $\fesca$ is anti-correlated with the dust-content\footnote{There is also little observational evidence for EW-boosting by a multiphase medium \citep[e.g.][]{Finkelstein08,Scarlata09}.} of galaxies \citep[][]{Atek09,Kornei10,Hayes11}. This correlation may explain why $\fesca$ increases with redshift from $\fesca \sim 1-3 \%$ at $z\sim 0$ to about $\fesca \sim 30-50\%$ at $z\sim 6$ \citep[e.g.][see Fig~\ref{fig:fesc}]{Hayes11,Blanc11,DJ13}, as the overall average dust content of galaxies decreases towards higher redshifts \citep[e.g.][]{Finkcolors,Bouwenscolor}. 
\begin{figure*}[t]
\vspace{-5mm}
\vbox{\centerline{\epsfig{file=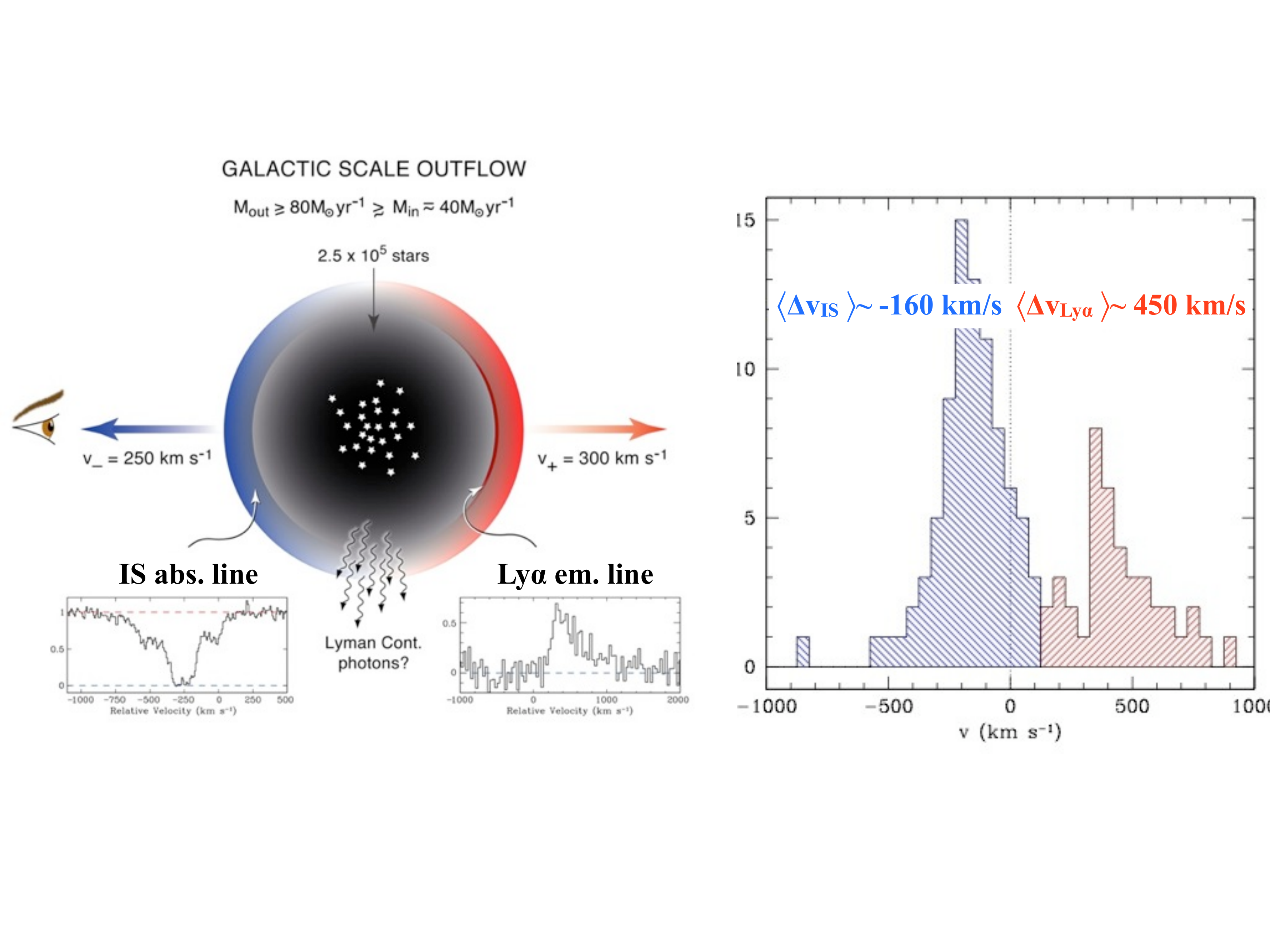,angle=0,width=12.5truecm}}}
\vspace{-20mm}
\caption[]{The Figure shows (some) observational evidence for the ubiquitous existence of cold gas in outflows in star-forming galaxies, and that this cold gas affects the Ly$\alpha$ transport: the {\it right panel} shows the vast majority of low-ionization interstellar (IS) absorption lines are blueshifted relative to the systemic velocity of the galaxy, which is indicative of outflows (as illustrated in the {\it left panel}, {\it Credit: Figure kindly provided by Max Pettini}). Moreover, the {\it right panel} ({\it Credit: from Figure~1 of Steidel et al. 2010 \textcopyright AAS. Reproduced with permission}) illustrates that the Ly$\alpha$ emission line is typically redshifted by an amount that is $\sim 2-3$ times larger than typical blueshift of the IS lines {\it in the same galaxies}. These observations are consistent with a scenario in which Ly$\alpha$ photons scatter back to the observer from the far-side of the nebular region (indicated schematically in the {\it left panel}).}
\label{fig:schemewind}
\end{figure*} 

It is worth cautioning here that observations are not directly constraining $\fesca$: Ly$\alpha$ photons that escape from galaxies can scatter frequently in the IGM (or circum-galactic medium) before reaching earth (see \S~\ref{sec:IGM}). This scattered radiation is typically too faint to be detected with direct observations \citep[][]{Zheng10}, and is effectively removed from observations. Stacking analyses, which must be performed with great caution \citep{Feldmeier}, have indeed revealed that there is increasing observational support for the presence of spatially extended Ly$\alpha$ halos around star-forming galaxies \citep[][but also see Jiang et al. 2013]{Fynbo99,Hayashino04,Hayes05,Hayes07,Rauch08,Ostlin09,Steidel10,Matsuda12,Hayes13,Momose14}. Direct observations of galaxies therefore measure the product\footnote{To differentiate the observationally inferred escape fraction from the real escape fraction, the former is referred to as the `effective' escape fraction, denoted with $\fesce$ \citep{Nagamine10,DJ13}.} of $\fesca$ and the fraction of photons that has {\it not} been scattered out of the line of sight.\\

To further understand the impact of RT one would ideally like to compare the properties of scattered Ly$\alpha$ photons (e.g. the spectrum) to that of nebular line photons that did not scatter, such as H$\alpha$ or [OIII]. For galaxies at $z>2$, these observations require spectrographs that operate in the NIR,  including e.g. NIRSPEC \citep{McL}, LUCIFER \citep{Seifert03}, and MOSFIRE \citep{McL2}.  

Simultaneous observations of Ly$\alpha$ \& other non-resonant nebular emission lines indicate that Ly$\alpha$ lines typically are redshifted with respect to these other lines by $\Delta v_{{\rm Ly}\alpha}$. This redshift is more prominent for the drop-out galaxies, in which the average $\Delta v_{{\rm Ly}\alpha}\sim 460$ km s$^{-1}$ in LBGs \citep{Steidel10,Kulas12}, which is larger than the shift observed in LAEs, where the average $\Delta v \sim 200$ km s$^{-1}$ \citep[][]{McLinden11,Chonis,Hashimoto13,McLinden14,Song14}\footnote{The different $\Delta v$ in LBGs and LAEs likely relates to the different physical properties of both samples of galaxies. \citet{Shibuya14} argue that LAEs may contain smaller $N_{\rm HI}$ which facilitates Ly$\alpha$ escape, and results in a smaller shift \citep[also see][]{Song14}.}. These observations indicate that outflows affect Ly$\alpha$ radiation while it is escaping from galaxies. This importance of outflows is not surprising: outflows are detected ubiquitously in absorption in other low-ionization transitions \citep[e.g.][]{Steidel10}. Moreover, the Ly$\alpha$ photons appear to interact with the outflow, as the Ly$\alpha$ line is redshifted by an amount that is correlated with the outflow velocity inferred from low-ionization absorption lines \citep[e.g.][]{Steidel10,Shibuya14}. Scattering of Ly$\alpha$ photons off these outflows provides the photons with a quick route to the wing of the line, where they can escape more easily. Indeed, earlier studies had indicated that gas kinematics plays a key role in the escape of Ly$\alpha$ photons from local galaxies \citep{Kunth98,Atek08}. The presence of winds and their impact on Ly$\alpha$ photons is illustrated explicitly in Figure~\ref{fig:schemewind}. \\

As modelling the outflowing component in interstellar medium is an extremely challenging task (see above), simple `shell-models' have been invoked to represent the scattering through outflows. In these shell-models, the outflow is represented by a spherical shell with a thickness that is 0.1$\times$ its inner/outer radius. The main properties that characterise the shell are its HI-column density, $N_{\rm HI}$, its outflow velocity, $v_{\rm sh}$, and its dust content \citep[e.g.][]{Ahn03,V06,V08}. For `typical' HI column densities in the range $N_{\rm HI}=10^{19}-10^{21}$ cm$^{-2}$ and $v_{\rm out}\sim$ a few hundred km s$^{-1}$, the red part of the spectrum peaks at $\sim 2v_{\rm out}$: photons that scatter 'back' to the observer on the far side of the Ly$\alpha$ source are Doppler boosted to twice the outflow velocity\footnote{Importantly, this argument implicitly assumes partially coherent back-scattering (see \S~\ref{sec:redist}).}, where they are sufficiently far in the wing of the absorption cross section to escape from the medium (the cross section at $\Delta v=200$ km s$^{-1}$ is only $\sigma_{\alpha}\sim$ a few times $10^{-20}$ cm$^2$, see Fig~\ref{fig:sigma}). The shell-model can reproduce observed Ly$\alpha$ spectral line shapes remarkably well \citep[e.g.][]{V08,SV08,Des10,V10}.

There are two issues with the shell-models though: ({\it i}) gas in the shells has a single outflow velocity and a small superimposed velocity dispersion, while observations of low-ionization absorption lines indicate that outflows typically cover a much wider range of velocities \citep[e.g.][]{Kulas12}; and ({\it ii}) observations of low-ionization absorption lines also suggest that outflows -while ubiquitous - do not completely surround UV-continuum emitting regions of galaxies. Observations by Jones et al. (2013) show that the maximum low-ionization covering fraction is $100\%$ in only 2 out of 8 of their $z>2$ galaxies (also see Heckman et al. 2011, who find evidence for a low covering factor of optically thick, neutral gas in a small fraction of lower redshift Lyman Break Analogues). There is thus some observational evidence that there exist sight lines that contain no detectable low-ionization (i.e. cold) gas, which may reflect the complex structure associated with outflows which cannot be captured with spherical shells. Two caveats are that ({\it a}) the inferred covering factors are measured as a function of velocity \citep[and can depend on spectral resolution, see e.g.][but Jones et al. 2013 discuss why this is likely not an issue in their analysis]{Pro06}. Gas at different velocities can cover different parts of the source, and the outflowing gas may still fully cover the UV emitting source. This velocity-dependent covering is nevertheless not captured by the shell-model; ({\it b}) the low-ionization metal absorption lines only probe enriched cold (outflowing) gas. Especially in younger galaxies it may be possible that there is additional cold (outflowing) gas that is not probed by metal absorption lines.

Recently, \citet{Shibuya14} have shown that Ly$\alpha$ line emission is stronger in galaxies in which the covering factor of low-ionization material is smaller (see their Fig~10). Similarly, Jones et al. (2012) found the average absorption line strength in low-ionization species to decrease with redshift, which again coincides with an overall increase in Ly$\alpha$ flux from these galaxies \citep{Stark10}. Besides dust, the covering factor of HI gas therefore plays an additional important role in the escape of Ly$\alpha$ photons. These cavities may correspond to regions that have been cleared of gas and dust by feedback processes \citep[see][who describe a simple `blow-out' model]{Nestor11,Nestor13}.\\

In short, dusty outflows appear to have an important impact on the interstellar Ly$\alpha$ radiative process, and give rise to redshifted Ly$\alpha$ lines. Low HI-column density holes further facilitate the escape of Ly$\alpha$ photons from the ISM, and can alter the emerging spectrum such that Ly$\alpha$ photons can emerge closer to the galaxies' systemic velocities \citep{Behrens14,Verhamme14}.

\subsection{Intergalactic Radiative Transfer}
\label{sec:IGM}
\begin{figure}[t]
\begin{center}
\includegraphics[width=7.5cm]{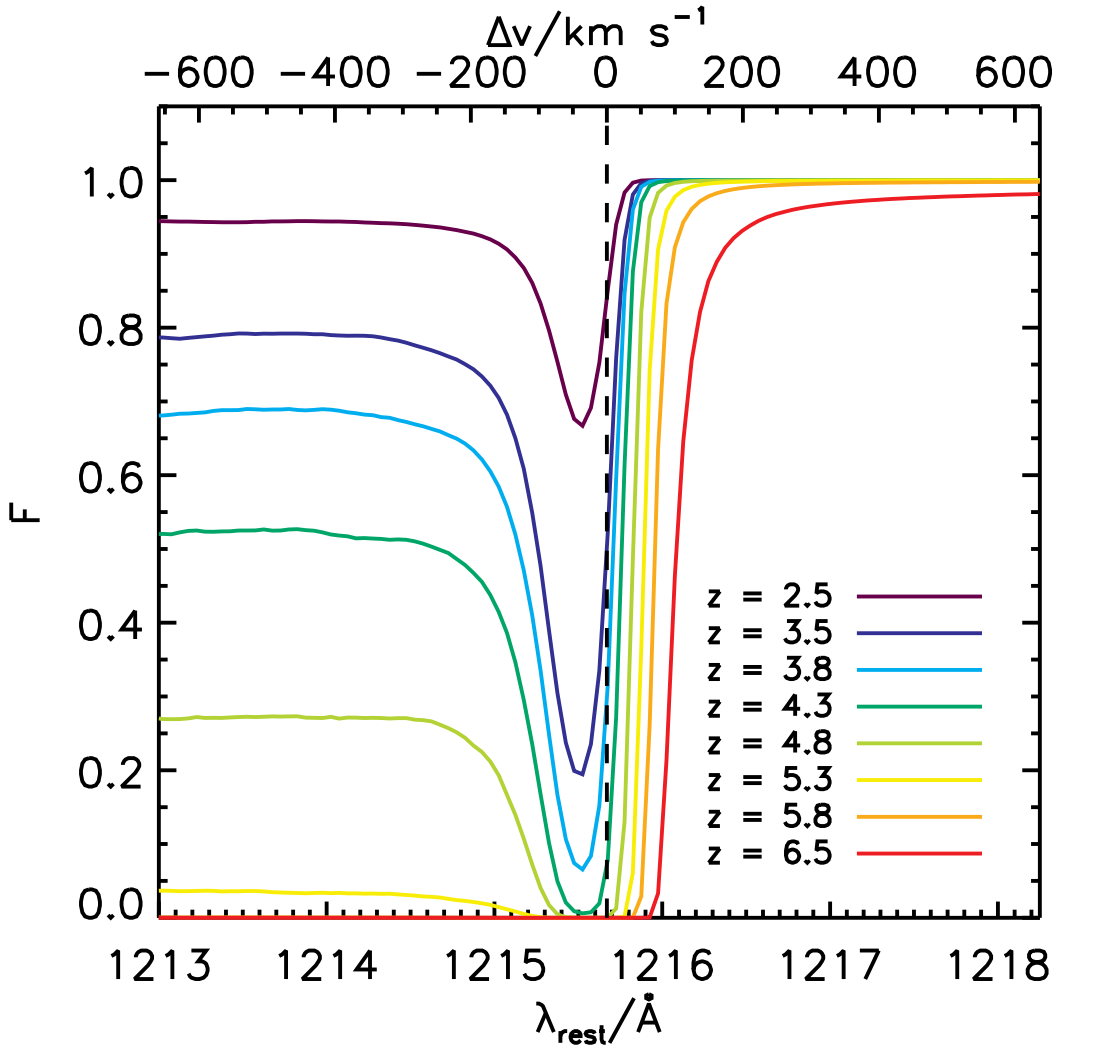}
\caption{The average fraction of photons that are transmitted though the IGM to the observer as a function of (restframe) wavelength. Each line represents a different redshift. At wavelengths well on the blue side of the line, we recover the mean transmission measured from the Ly$\alpha$ forest. Overdense gas at close proximity to the galaxy increases the IGM opacity close to the Ly$\alpha$ resonance (and causes a dip in the transmission curve, {\it Credit: from Figure~2 of Laursen et al. 2011 \textcopyright AAS. Reproduced with permission}).}
\label{fig:IGM}
\end{center}
\end{figure}
As mentioned above, to understand the impact of reionization on Ly$\alpha$ emitting galaxies, it is important to also understand the impact of the IGM when it has been fully reionized. The transmitted fraction of photons on the blue-side of the Ly$\alpha$ resonance relates to the Gunn-Peterson optical depth \citep[][]{GP65},
\begin{equation}
\tau_{\rm GP}(\nu_{\rm em})=\int_0^{\infty}ds \hs n_{\rm HI}(s)\sigma_{\alpha}(\nu[s,\nu_{\rm em}]),
\end{equation} where the line integral starts at the galaxy. As photons propagate a proper differential distance $ds$, the cosmological expansion of the Universe redshifts the photons by an amount $d\nu = -ds H(z) \nu/c$. Photons that were initially emitted at $\nu_{\rm em} >\nu_{\alpha}$ will thus redshift into the line resonance. Because $\sigma_{\alpha}(\nu)$ is peaked sharply around $\nu_{\alpha}$ (see Fig~\ref{fig:sigma}), we can approximate this integral by taking $n_{\rm HI}(s)$ and $c/\nu\approx \lambda_{\alpha}$ outside of the integral. If we further assume that $n_{\rm HI}(s)$ corresponds to $\bar{n}_{\rm HI}(z)$ - where $\bar{n}_{\rm HI}(z)=\Omega_{\rm b}h^2(1-Y_{\rm He})(1+z)^3/m_{\rm p}$ denotes the mean number density of hydrogen atoms in the Universe at redshift $z$ - then we obtain

\begin{align}
\tau_{\rm GP}(\nu_{\rm em} > \nu_{\alpha})=\frac{\bar{n}_{\rm HI}(z)\lambda_{\alpha}}{H(z)}\int_0^{\infty}d\nu \hs \sigma_{\alpha}(\nu) = \\ \nonumber
=\frac{\bar{n}_{\rm HI}(z)\lambda_{\alpha}}{H(z)}f_{\alpha}\frac{\pi e^2}{m_{\rm e}c} \approx 7.0 \times 10^5 \Big{(} \frac{1+z}{10}\Big{)}^{3/2},
\end{align} where we used that $\int d\nu\hs\sigma(\nu)=f_{\alpha}\frac{\pi e^2}{m_{\rm e}c}$ \citep[e.g.][p 102]{RL79}. Quasar absorption line spectra indicate that the IGM transmits an {\it average} fraction $F \sim 85\%$ [$F \sim 40\%$] of Ly$\alpha$ photons at $z=2$ [$z=4$] which imply `effective optical depths of $\tau_{\rm eff} \equiv -{\rm ln}[F]  \sim 0.15$ [$\tau_{\rm eff} \sim 0.9$] \citep[e.g.][]{FG08}. The measured values $\tau_{\rm eff} \ll \tau_{\rm GP}$ which is (of course) because the Universe was highly ionized at these redshifts. A common approach to model the impact of the IGM is to reduce the Ly$\alpha$ flux on the blue side of the Ly$\alpha$ resonance by this observed (average) amount, while transmitting all flux on the red side.\\

The measured values of $F$ and $\tau_{\rm eff}$ are averaged over a range of frequencies. In detail, density fluctuations in the IGM give rise to enhanced absorption in overdense regions which is observed as the Ly$\alpha$ forest.  It is important to stress that galaxies populate overdense regions of the Universe in which: ({\it i}) the gas density was likely higher than average, ({\it ii}) peculiar motions of gas attracted by the gravitational potential of dark matter halos change the relation between $ds$ and $d\nu$, ({\it iii}) the local ionising background was likely elevated. We thus clearly expect the impact of the IGM\footnote{Early studies defined the IGM to be all gas at $r>1-1.5$ virial radii, which would correspond to the `circum-galactic' medium by more recent terminology. Regardless of what we call this gas, scattering of Ly$\alpha$ photons would remove photons from a spectrum of a galaxy, and redistribute these photons over faint, spatially extended Ly$\alpha$ halos.} at frequencies close to the Ly$\alpha$ emission line to differ from the mean transmission in the Ly$\alpha$ forest: Figure~\ref{fig:IGM} shows the transmitted fraction of Ly$\alpha$ photons averaged over a large number of sight lines to galaxies in a cosmological hydrodynamical simulation (Laursen et al. 2011). This Figure shows that  infall of over dense gas (and/or retarded Hubble flows) around dark matter halos hosting Ly$\alpha$ emitting galaxies can give rise to an increased opacity of the IGM around the Ly$\alpha$ resonance, and even extending somewhat into the red side of the Ly$\alpha$ line \citep[][]{Santos04,IGM,Laursen11}. We denote the opacity of the ionized IGM at velocity off-set $\Delta v$ and redshift $z$ with $\tauHII(z,\Delta v)$. This provides a source of intergalactic opacity additional to the `damping wing' optical depth, $\taud(z,\Delta v)$, that is only relevant during reionization (see \S~\ref{sec:eorlya}).
\begin{figure*}[t]
\vbox{\centerline{\epsfig{file=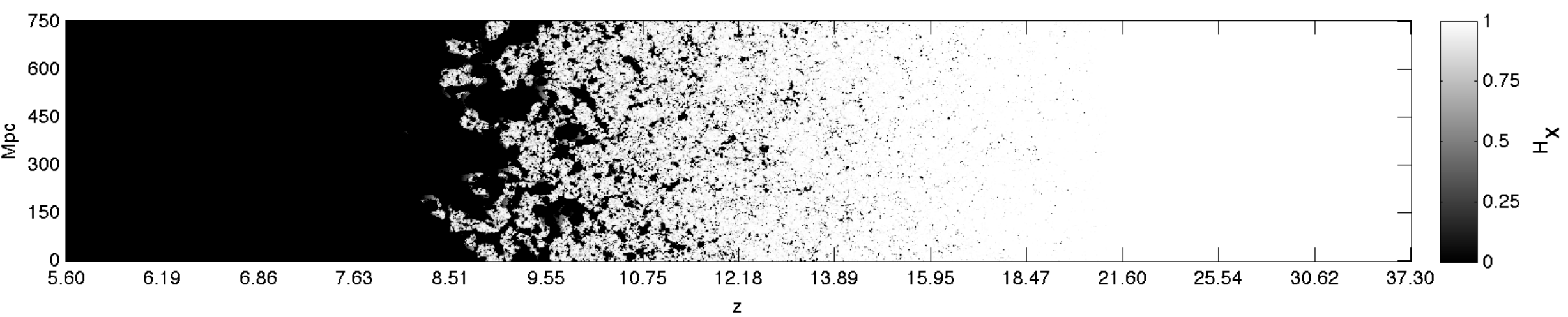,angle=0,width=17.0truecm}}}
\vspace{0mm}
\caption[]{The predicted redshift evolution of the ionization state of the IGM in a realistic reionization model ({\it Credit: Figure kindly provided by Andrei Mesinger}). The white/black represents fully neutral/ionized intergalactic gas. This Figure demonstrates the inhomogeneous nature of the reionization process which took places over an extended range of redshifts: at $z >16$ the first ionized regions formed around the most massive galaxies in the Universe (at that time). During the final stages of reionization - here at $z\sim 9$ the IGM contains ionized bubbles several tens of cMpc across.}
\label{fig:reionization}
\end{figure*} 

Because these models predict that the IGM can strongly affect frequencies close to the Ly$\alpha$ resonance, the overall impact of the IGM depends strongly on the Ly$\alpha$ spectral line shape as it emerges from the galaxy. For Gaussian and/or generally symmetric emission lines centered on the galaxies' systemic velocities, the IGM can transmit as little as\footnote{It is worth noting that these models predict that the IGM can reduce the observed Ly$\alpha$ line by as much as $\sim 30\%$ between $z=5.7$ and $z=6.5$ \citep[][]{Laursen11}.} $\mathcal{T}_{\rm IGM}=10-30\%$ even for a fully ionized IGM \citep[e.g.][]{IGM,Zheng10,Dayal11,Laursen11}. However, when scattering through outflows shifts the line sufficiently away from line centre, the overall impact of the IGM can be reduced tremendously \citep[e.g.][]{Haiman02,Santos04,D11,Garel}.  

We stress that the IGM opacity discussed above originates in mildly over dense ($\delta=1-20$, see Dijkstra et al. 2007a), highly ionized gas.  Another source of opacity is provided by Lyman-limit systems (LLSs) and Dampled Ly$\alpha$ absorbers (DLAs). The abundance of LLSs increases (weakly) with redshift at $z>3$ \citep[e.g.][]{Rahmati13}, and may strongly affect the IGM opacity at $z>6$ \citep[][but see Mesinger et al. 2014]{Bolton}.

\section{Reionization and the Visibility of the Ly$\alpha$ Line}
\label{sec:EoR}

\subsection{Reionization}
Reionization refers to the process during which intergalactic gas was transformed from fully neutral to fully ionized. For reviews on the Epoch of Reionization (EoR) we refer the reader to e.g. \citet{BL01}, \citet{Frev}, and \citet{Morales10}. The EoR is characterized by the existence of a patches of diffuse neutral intergalactic gas, which provides a source of opacity to Ly$\alpha$ photons additional to that discussed in \S~\ref{sec:IGM}. 

Reionization was likely not a homogeneous process in which the ionization state of intergalactic hydrogen changed everywhere in the Universe at the same time, and at the same rate. Instead, it was likely an extended highly inhomogeneous process. The first sources of ionizing radiation were highly biased tracers of the underlying mass distribution. As a result, these galaxies were clustered on scales of tens of comoving Mpc (cMpc, Barkana \& Loeb 2004). The strong clustering of these first galaxies in overdense regions of the Universe caused these regions to be reionized first \citep[e.g. Fig~ 1 of][]{WL07}, which thus created fluctuations in the ionization field over similarly large scales. As a result a proper description of the reionization process requires simulations that are at least 100 cMpc on the side \citep[e.g.][]{Trac11}. 

Ideally, one would like to simulate reionization by performing full radiative transfer calculations of ionising photons on cosmological hydrodynamical simulations. A number of groups have developed codes that can perform these calculations in 3D \citep[e.g.][]{Gnedin00,Sokasian,Ciardi03,Mellema06,Trac07,PS08,Finlator09}. These calculations are computationally challenging as one likes to simultaneously capture the large scale distribution of HII bubbles, while resolving the photon sinks (such as Lyman Limit systems) and the lowest mass halos ($M\sim 10^8$ M$_{\odot}$) which can contribute to the ionising photon budget \citep[see e.g.][]{Trac11}. Modeling reionization contains many poorly known parameters related to galaxy formation, the ionising emissivity of star-forming galaxies, their spectra etc.  Alternative, faster `semi-numeric' algorithms have been developed which allow for a more efficient exploration of the full the parameter space \citep[e.g.][]{DexM}. These semi-numeric algorithms utilize excursion-set theory to determine if a cell inside a simulation is ionized or not \citep[][]{Furlanetto04}. Detailed comparisons between full radiation transfer simulations and semi-numeric simulations show both methods produce very similar ionization fields \citep{Zahn11}.

The picture of reionization that has emerged from analytical consideration and large-scale simulations is one in which the early stages of reionization are characterized by the presence of HII bubbles centered on overdense regions of the Universe, completely separated from each other by a neutral IGM \citep{Furlanetto04,Iliev06,McQuinn07}. The ionized bubbles grew in time, driven by a steadily growing number of star-forming galaxies residing inside them. The final stages of reionization are characterized by the presence of large bubbles, whose individual sizes exceeded tens of cMpc (e.g. Zahn et al. 2011). Ultimately these bubbles overlapped (percolated), which completed the reionization process. The predicted redshift evolution of the ionization state of the IGM in a realistic reionization model is shown in Figure~\ref{fig:reionization}. This Figure nicely illustrates the inhomogeneous, temporally extended nature of the reionization process.

\subsection{Inhomogeneous Reionization \& Its Impact on Ly$\alpha$}
\label{sec:eorlya}
\begin{figure*}
\vbox{\centerline{\epsfig{file=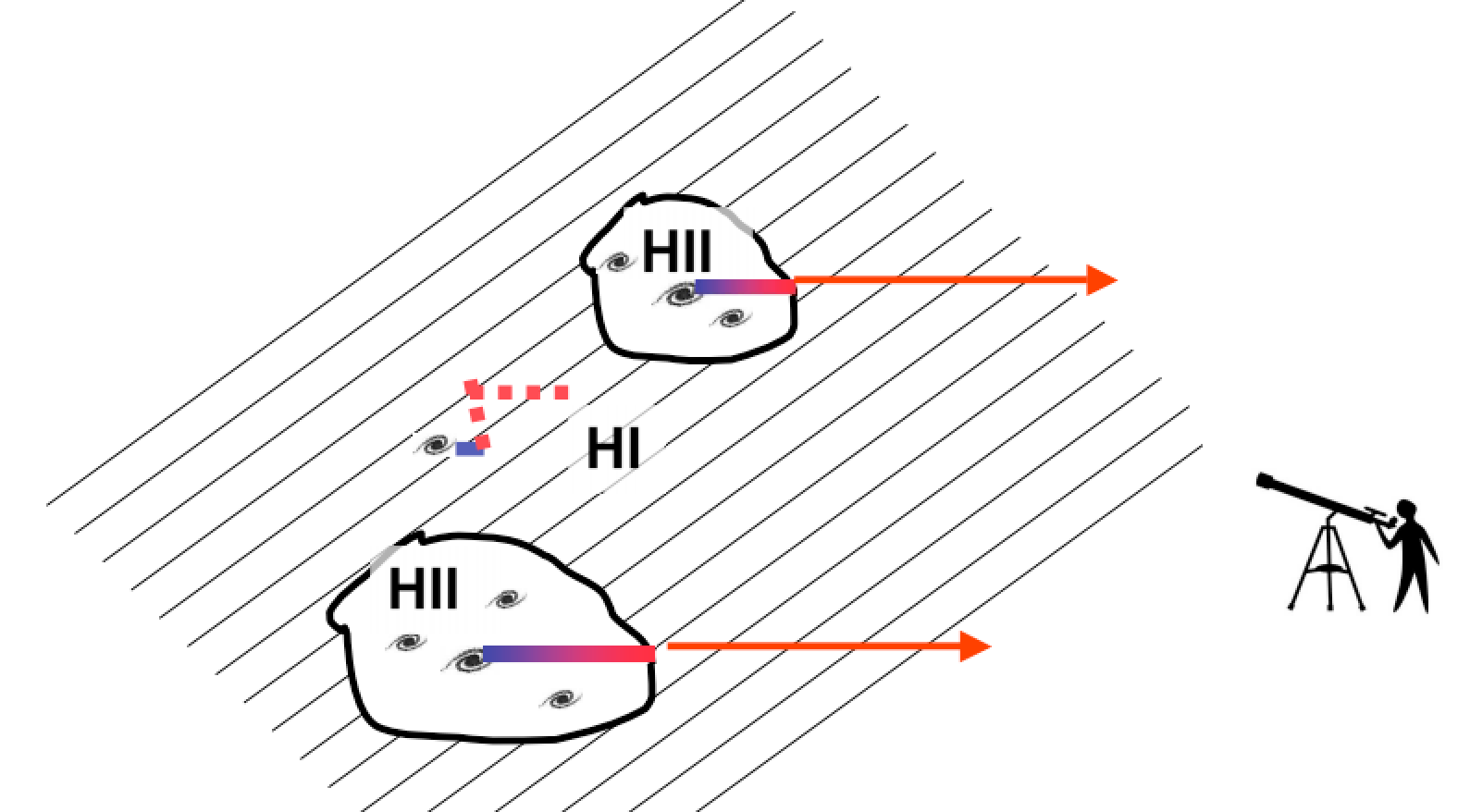,angle=0,width=10.0truecm}}}
\vspace{0mm}
\caption[]{This Figure schematically shows why inhomogeneous reionization boosts the visibility of Ly$\alpha$ emitting galaxies. During the mid and late stages of reionization star-forming - and hence Ly$\alpha$ emitting - galaxies typically reside in large HII bubbles. Ly$\alpha$ photons emitted inside these HII bubbles can propagate - and redshift away from line resonance - through the ionized IGM before encountering the neutral IGM. The resulting reduced opacity of the neutral IGM (Eq~\ref{eq:gp3}) to Ly$\alpha$ photons enhances the prospect for detecting Ly$\alpha$ emission from those galaxies inside HII bubbles.}
\label{fig:schemeEoR}
\end{figure*} 
Star-forming galaxies that are luminous enough to be detected with existing telescopes likely populated dark matter halos with masses in excess of $M \gsim 10^{10}$ M$_{\odot}$.  These halos preferentially resides in over dense regions of the Universe, which were reionized earliest. It is therefore likely that (Ly$\alpha$ emitting) galaxies preferentially resided inside these large HII bubbles. Ly$\alpha$ photons emitted by galaxies located inside these HII regions can propagate - and therefore redshift away from line resonance - through the ionized IGM before encountering the neutral IGM. Because of the strong frequency-dependence of the Ly$\alpha$ absorption cross section, these photons are less likely to be scattered out of the line of sight. A non-negligible fraction of Ly$\alpha$ photons may be transmitted directly to the observer, which is illustrated schematically in Figure~(\ref{fig:schemeEoR}). Inhomogeneous reionization thus enhances the prospect for detecting Ly$\alpha$ emission from galaxies inside HII bubbles \citep[e.g.][]{Madau00,Haiman02,Gnedin04,Furlanetto04,Furlanetto06,McQuinn07,Mesinger08,Iliev08,Dayal11,D11,Jensen,Hutter14}. This Figure also illustrates that Ly$\alpha$ photons emitted by galaxies that lie outside of large HII bubbles, scatter repeatedly in the IGM. These photons diffuse outward, and are visible only faint extended Ly$\alpha$ halos \citep[][]{LR99,Kobayashi06,JD12}. \\
 
We can quantify the impact of neutral intergalactic gas on the Ly$\alpha$ flux from galaxies following our analysis in \S~\ref{sec:IGM}. We denote the optical depth in the neutral intergalactic patches with $\tau_{\rm D}$. We first consider the simplest case in which a Ly$\alpha$ photon encounters one fully neutral patch which spans the line-of-sight coordinate from $s_b$ (`b' stands for beginning) to $s_e$ (`e' stands for end):
\begin{equation}
\tau_{\rm D}(\nu)=\int_{s_b}^{s_e}ds\hs n_{\rm HI}(s)\sigma_{\alpha}(\nu[s]).
\label{eq:1patch}
\end{equation} Following the analysis of \S~\ref{sec:IGM} we change to frequency variables, and recast Eq~\ref{eq:1patch} as
\begin{equation}
\tau_{\rm D}(\nu)=\tau_{\rm GP}\frac{\int_{\nu_b(\nu)}^{\nu_e(\nu)}d\nu' \sigma_{\alpha}(\nu')}{\int_0^{\infty}d\nu' \sigma_{\alpha}(\nu')}.
\label{eq:1patchb}
\end{equation} 
The denominator can be viewed as a normalisation constant, and we can rewrite Eq~\ref{eq:1patchb} as
\begin{equation}
\tau_{\rm D}(\nu)=\frac{\tau_{\rm GP}}{\sqrt{\pi}} \int_{x_b(\nu)}^{x_e(\nu)} dx\hs \phi(x'),
\label{eq:1patchc}
\end{equation} where the factor of $\sqrt{\pi}$ enters because of our adopted normalisation for the Voigt profile $\phi(x)$.\\

During the EoR a Ly$\alpha$ photon emitted by a galaxy will generally propagate through regions that are alternating between (partially) neutral and highly ionized. The more general case should therefore contain the sum of the optical depth in separate neutral patches. 
\begin{equation}
\tau_{\rm D}(\nu)=\frac{1}{\sqrt{\pi}}\sum_i \tau_{\rm GP,i} \hs x_{\rm HI,i}\int_{x_{b,i}(\nu)}^{x_{e,i}(\nu)} dx\hs \phi(x'),
\label{eq:morepatch}
\end{equation} where we have placed $\tau_{\rm GP}$ within the sum, because $\tau_{\rm GP}$ depends on redshift as $\tau_{\rm GP} \propto (1+z_{\rm i})^{3/2}$, and therefore differs slightly for each neutral patch (at redshift $z_{\rm i}$).

More specifically,  the total optical depth of the neutral IGM to Ly$\alpha$ photons emitted by a galaxy at redshift $z_{\rm g}$ with some velocity off-set $\Delta v$ is given by Eq~\ref{eq:morepatch} with $x_{b,i} = \frac{-1}{v_{\rm th,i}}[\Delta v+H(z_{\rm g})R_{b,i}/(1+z_{\rm g})]$, in which $R_{b,i}$ denotes the comoving distance to the beginning of patch `i' ($x_{e,i}$ is defined similarly). Eq~\ref{eq:morepatch} must generally be evaluated numerically. However, one can find intuitive approximations: for example, if we assume that ({\it i}) $x_{\rm HI,i}=1$ for all `i', ({\it ii})  $z_{\rm i} \sim z_{\rm g}$, and ({\it iii}) that Ly$\alpha$ photons have redshifted away from resonance by the time they encounter this first neutral patch\footnote{If a photon enters the first neutral patch on the blue side of the line resonance, then the total opacity of the IGM depends on whether the photon redshifted into resonance inside or outside of a neutral patch. If the photon redshifted into resonance inside patch `i', then $\tau_{\rm D}(z_{\rm g},\Delta v)=\tau_{\rm GP}(z)x_{\rm HI,i}$. If on the other hand the photon redshifted into resonance in an ionized bubble, then we must compute the optical depth in the ionized patch, $\tauHII(z,\Delta v=0)$, plus the opacity due to subsequent neutral patches. Given that the ionized IGM at $z=6.5$ was opaque enough to completely suppress Ly$\alpha$ flux on the blue-side of the line, the same likely occurs inside ionized HII bubbles during reionization because of ({\it i}) the higher intergalactic gas density, and ({\it ii}) the shorter mean free path of ionizing photons and therefore likely reduced ionizing background that permeates ionised HII bubbles at higher redshifts.}, then
\begin{equation}
\tau_{\rm D}(z_{\rm g},\Delta v)=\frac{\tau_{\rm GP}(z_{\rm g})}{\sqrt{\pi}}\sum_i \hs \Big{(}\frac{a_{\rm v,i}}{\sqrt{\pi}x_{e,i}} - \frac{a_{\rm v,i}}{\sqrt{\pi}x_{b,i}}\Big{)},
\label{eq:morepatchII} 
\end{equation} where $x_{e,i}=x_{e,i}(\Delta v)$ and  $x_{b,i}=x_{b,i}(\Delta v)$. It is useful to explicitly highlight the sign-convention here: photons that emerge redward of the Ly$\alpha$ resonance have $\Delta v > 0$, which corresponds to a negative $x$. Cosmological expansion redshifts photons further, which decreases $x$ further. The $\av/[\sqrt{\pi}x_{b,i}]$ is therefore more negative, and $\tau_{\rm D}$ is thus positive. 

One can define the `patch-averaged' neutral fraction  $\bar{x}_{\rm D}$ - which is related to the volume filling factor of neutral hydrogen $\langle x_{\rm HI} \rangle$ in a non-trivial way (see Mesinger \& Furlanetto 2008) - as
\begin{align}
\tau_{\rm D}(z_{\rm g},\Delta v)=\frac{\tau_{\rm GP}(z_{\rm g})}{\sqrt{\pi}}\bar{x}_{\rm D}\Big{(}\frac{\av}{\sqrt{\pi}x_{e}} - \frac{\av}{\sqrt{\pi}x_{b,1}}\Big{)} \approx \\ \nonumber
\approx \frac{\tau_{\rm GP}(z_{\rm g})}{\pi}\bar{x}_{\rm D} \frac{\av}{|x_{b,1}|}=\frac{\tau_{\rm GP}(z_{\rm g})}{\pi}\bar{x}_{\rm D}\frac{A_{\alpha}c}{ 4\pi \nu_{\alpha}}\frac{1}{\Delta v_{b,1}}, 
\end{align} where $x_{\rm e}$ denotes the frequency that photon has redshifted to when it exits from the last neutral patch, while $x_{b,1}$ denotes the photon's frequency when it encounters the first neutral patch. Because typically $|x_{\rm e}|\gg |x_{b,1}|$ we can drop the term that includes $x_{\rm e}$. We further substituted the definition of the Voigt parameter $\av =A_{\alpha} /(4 \pi \Delta \nu_{\rm D})$, to rewrite $x_{b,1}$ as a velocity off-set from line resonance when a photon first enters a neutral patch, $\Delta v_{b,1} = \Delta v+H(z_{\rm g})R_{b,i}/(1+z_{\rm g})$. 

Substituting numbers gives \citep[e.g.][]{M98,DW10}

\begin{figure*}
\centerline{\includegraphics[width=65mm]{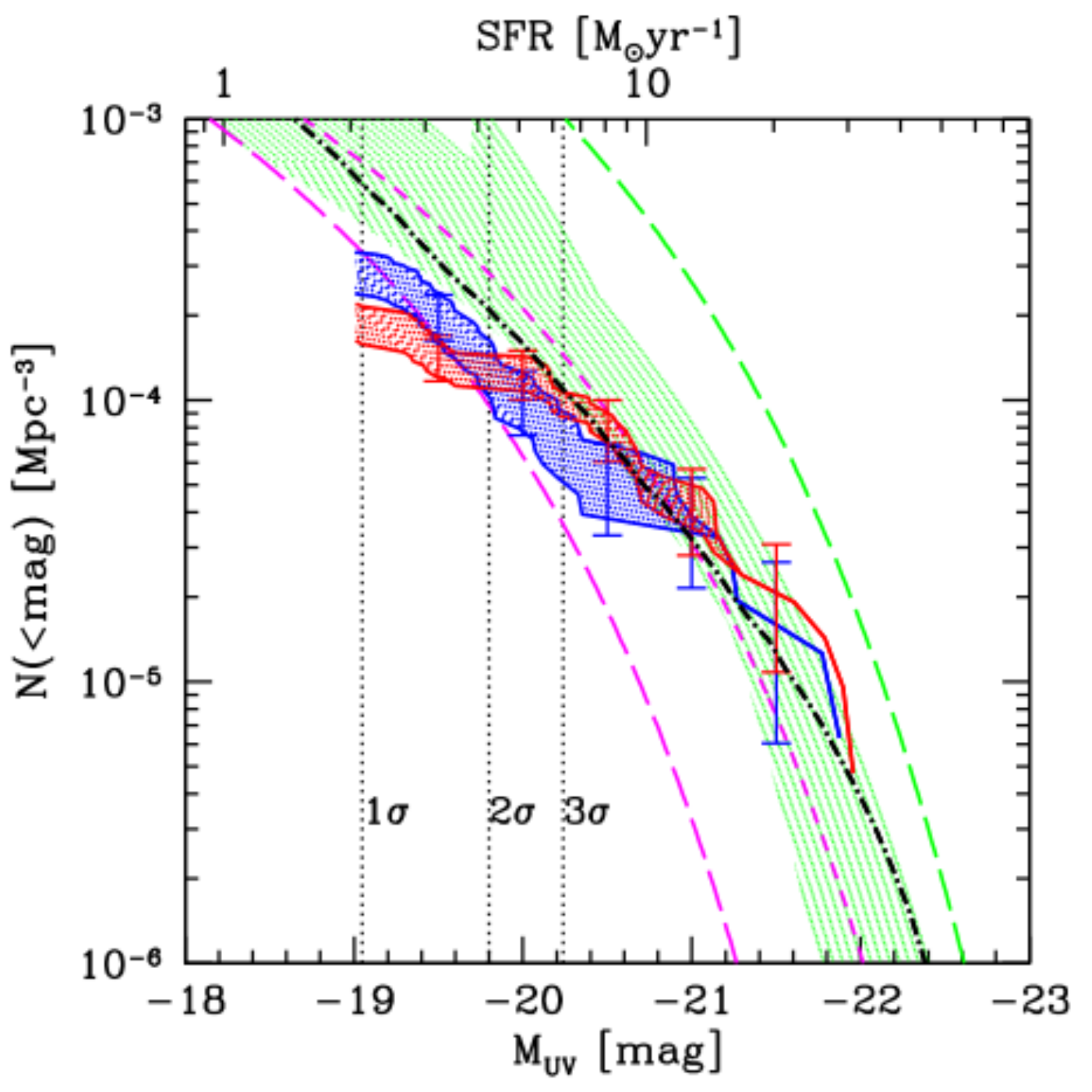}
\includegraphics[width=65mm]{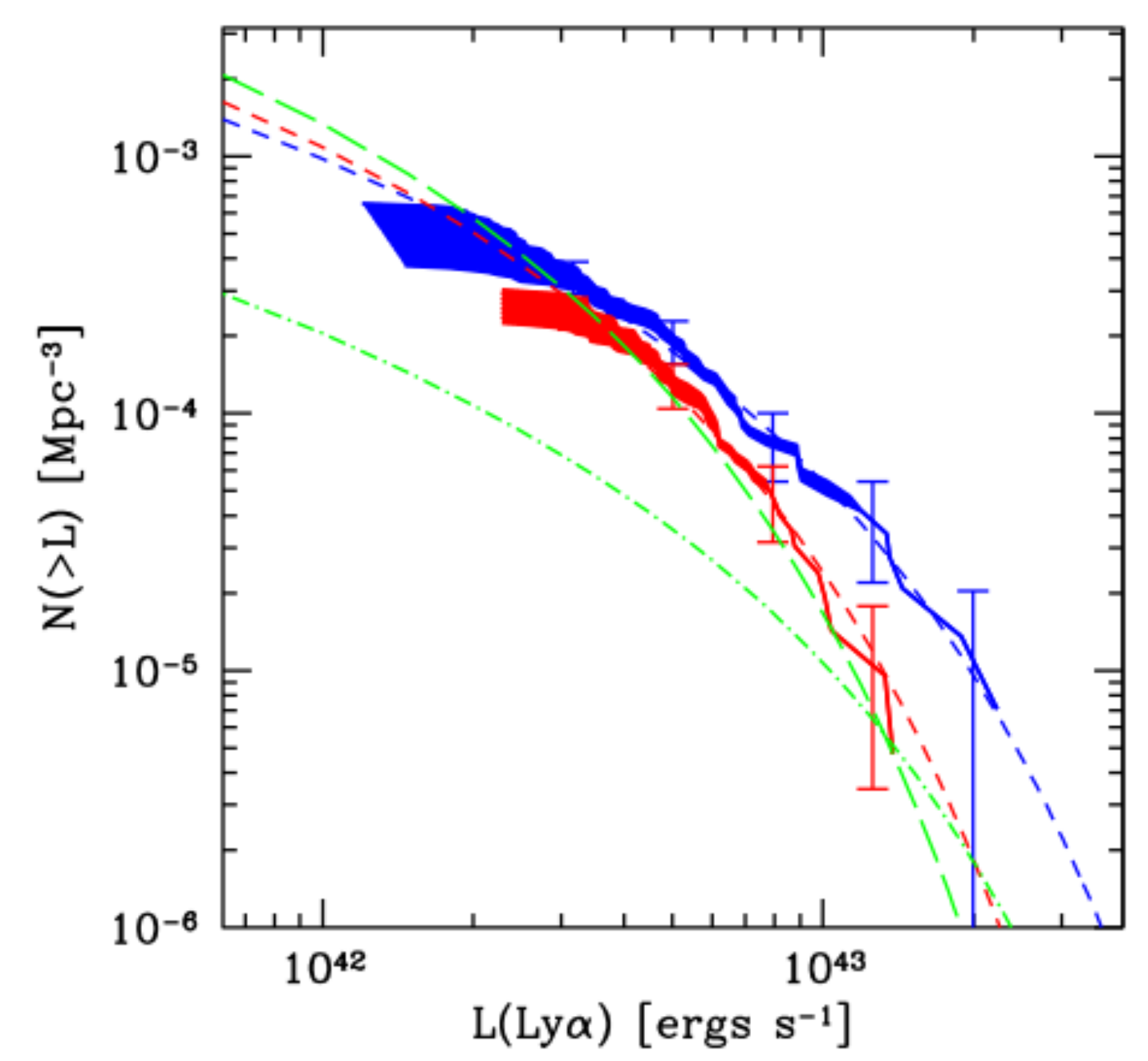}}
\caption[]{Redshift evolution of LAE luminosity functions (LFs) between $z=5.7$ and $z=6.5$. The {\it left}/{\it right panel} shows the UV/Ly$\alpha$ LFs. {\it Blue/red solid lines} correspond to the acceptable ranges of the luminosity functions. {\it Blue/red dashes lines} show best fit Schechter functions assuming a faint end slope $\alpha=-1.5$. The reduction in the Ly$\alpha$ luminosity function at $z>5.7$ ({\it right panel}) differs from the observed non-evolution between $z=3.1-5.7$. Importantly, the reduction in the Ly$\alpha$ LF appears to arise from a reduction in the Ly$\alpha$ flux, as the UV-continuum LF of LAEs does not evolve between $z=5.7$ and $z=6.5$ ({\it Credit: from Figure~7 and Figure~9 of Kashikawa et al. 2011 \textcopyright AAS. Reproduced with permission}), we refer the reader to that paper for a full description of all other lines).}
\label{fig:LAELF}
\end{figure*} 
\begin{equation}
\tau_{\rm D}(z_{\rm g},\Delta v)\approx 2.3\bar{x}_{\rm D} \Big{(} \frac{\Delta v_{b,1}}{600\hs{\rm km}\hs{\rm s}^{-1}}\Big{)}^{-1}\Big{(} \frac{1+z_{\rm g}}{10}\Big{)}^{3/2}. 
 \label{eq:gp3}
\end{equation}
This equation shows that the opacity of the IGM drops dramatically once photons enter the first patch of neutral IGM with a redshift. This redshift may occur partly at the interstellar level, and partly at the intergalactic level: scattering off outflowing material\footnote{Scattering through an extremely opaque static medium gives rise to a double-peaked Ly$\alpha$ spectrum (see Fig~\ref{fig:static}). Of course, photons in the red peak start with a redshift as well, which boosts their visibility especially for large $N_{\rm HI}$ \citep[see Fig~2 in][also see Haiman 2002]{DW10}.} at the interstellar level can efficiently redshift Ly$\alpha$ photons by a few hundred km/s (see \S~\ref{sec:ISM}).  Because Ly$\alpha$ photons can undergo a larger cosmological subsequent redshift inside larger HII bubbles, Ly$\alpha$ emitting galaxies inside larger HII bubbles may be more easily detected\footnote{Eq~\ref{eq:gp3} shows that setting $\tau_{\rm D}=1$ for $\bar{x}_{\rm D}=x_{\rm HI}=1.0$ requires $\Delta v = 1380$ km s$^{-1}$. This cosmological redshift requires the nearest patch to be $d_{\rm patch}\sim 1$ Mpc (proper) away. This required distance is independent of $z$, because at a fixed $d_{\rm patch}$, the corresponding cosmological redshift $\Delta v \propto (1+z)^{3/2}$ \citep{M98}.}. The presence of large HII bubbles during inhomogeneous reionization may have drastic implications for the prospects of detecting Ly$\alpha$ emission from the epoch.

\section{Observations \& Interpretation}
\label{sec:obsint}

\subsection{The Data}
\label{sec:obs}

There is increasing observational support for the notion that Ly$\alpha$ emission from star-forming galaxies becomes stronger out to $z=6$, after which it suddenly becomes weaker. This applies to both LAEs and drop-out galaxies:
\begin{itemize}[leftmargin=*]
\item Observations of Ly$\alpha$ selected galaxies, also known as Ly$\alpha$ emitters (LAEs), have indicated that the Ly$\alpha$ luminosity functions (LFs) of LAEs do not evolve between $z=3.1$ and $z=5.7$ \citep[e.g.][]{Hu98,Ouchi08}. In contrast, the Ly$\alpha$ luminosity of LAEs does evolve at $z>6$ \citep[e.g.][]{K06,Ouchi10,Ota10,K11}, as is shown in the {\it right panel} of Figure~\ref{fig:LAELF}. Importantly, the UV luminosity function of LAEs does not evolve between $z=5.7$ and $z=6.5$ (see the {\it left panel} of Fig~\ref{fig:LAELF}), which suggests that the reduction of the Ly$\alpha$ luminosity functions at $z>6$ is a result of a reduction in Ly$\alpha$ flux coming from these galaxies \citep[as opposed to a reduction simply in their number density,][]{K06,K11}. This evolution appears to continue towards at higher redshifts ($z\gsim 7$) as well, where Ly$\alpha$ detections are rare \citep[][but also see Tilvi et al. 2010]{Ota10,Shibuya12,Konno14}, often yielding null-detections \citep[e.g.][]{Clement12,Ota12,Jiang13b,Faisst14}.

\begin{figure}
\vbox{\centerline{\epsfig{file=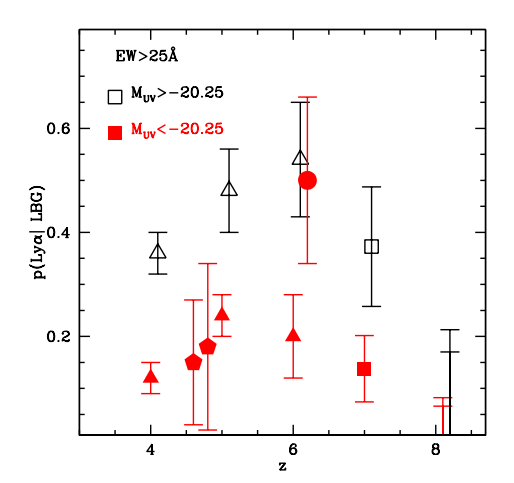,angle=0,width=8.0truecm}}}
\caption[]{Redshift evolution of the fraction of drop-out galaxies with `strong' Ly$\alpha$ emission (EW $>$ 25 \AA). This fraction, also known as the `Ly$\alpha$' fraction, increases with redshift out to $z=3$, but then decreases ({\it Credit: from Figure~4 of Treu et al. 2013 \textcopyright AAS. Reproduced with permission}). This redshift evolution is quantitatively consistent with the redshift evolution of Ly$\alpha$ luminosity functions of LAEs. }
\label{fig:Lyafraction}
\end{figure} 

\item A similar trend has been observed in drop-out selected galaxies (or Lyman Break Galaxies, or LBGs)\footnote{Formally LBGs correspond to galaxies selected via the drop-out technique at $z=2-4$. At lower redshift these galaxies would be BX, or BzK galaxies. At higher redshifts, these galaxies are often simply referred to as 'drop-outs'.}
: The fraction of drop-out galaxies with `strong' Ly$\alpha$ emission lines increases within this redshift range \citep[][]{Stark10}, which can quantitatively account for the observed non-evolution of the Ly$\alpha$ LFs within that same redshift range \citep[see e.g.][]{DW12}. However, at $z>6$ the Ly$\alpha$ fraction decreases \citep[][]{Fontana10,Pentericci,Schenker,Ono,Ca12,Ca13,Finkelstein13,P14,Schenker14,Tilvi14}, in line with observations of  the Ly$\alpha$ and UV luminosity functions of LAEs (again quantitatively consistent, see Dijkstra et al. 2014). This observed evolution is shown in Figure~\ref{fig:Lyafraction}.

\end{itemize}
It is therefore safe to conclude that observations of galaxies indicate that {\it the Ly$\alpha$ line has greater difficulty reaching us from galaxies at $z>6$ than what expected from observations at $z<6$}. This is a remarkable result, especially since it has long been predicted to be a key signature of reionization.

\begin{figure*}
\vspace{-60mm}
\vbox{\centerline{\epsfig{file=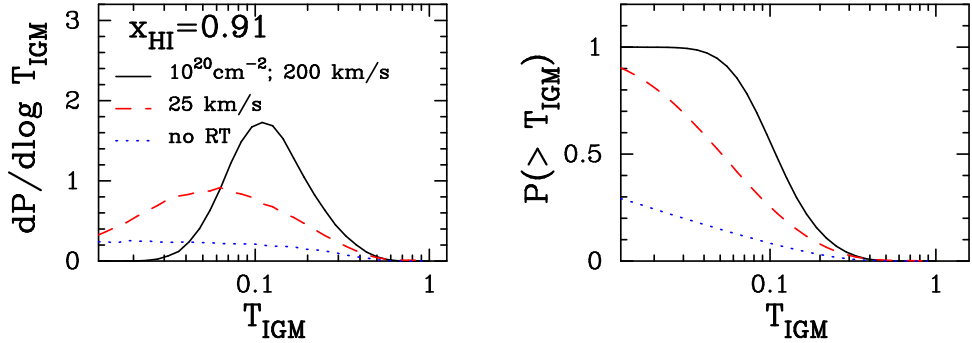,angle=0,width=16.0truecm}}}
\vspace{-10mm}
\caption[]{These Figures show the probability distributions ({\it left panel}: differential, {\it right panel}: cumulative) of the fraction of photons that are transmitted through a model for the IGM whose volume averaged neutral fraction is $\langle x_{\rm HI} \rangle =91\%$. The {\it blue dotted line} represents a realistic inhomogeneous reionization model, in which Ly$\alpha$ photons escape from galaxies at their systemic velocities. The {\it red dashed line} \& {\it black solid line} show models in which the Ly$\alpha$ photons emerge from galaxies with a redshift as a result from scattering through an optically thick outflow (see text for details). This plot shows that inhomogeneous reionization allows $>10\%$ of Ly$\alpha$ photons to be transmitted through a highly neutral IGM (which would correspond to an 'effective' optical depth $\langle \tau_{\rm GP} \rangle<$2.3) for a non-negligible fraction of galaxies, and that this fraction goes up dramatically when winds affect the Ly$\alpha$ photons on interstellar scales ({\it Credit: from Figure~2 of Dijkstra et al. 2011, `The detectability of Ly$\alpha$ emission from galaxies during the epoch of reionization', MNRAS, 414, 2139D.}).}
\label{fig:TIGM}
\end{figure*} 

\subsection{Interpretation of the Data}
\label{sec:interpretation}
The sudden reduction in Ly$\alpha$ flux beyond $z\sim 6$ is naturally associated with the emergence of neutral patches of intergalactic gas. It is thus natural to ask: {\it if} this reduction is indeed due to the presence of neutral patches of IGM during the EoR, {\it then} how does this reduction compare to what is expected in the context of inhomogeneous reionization?

The fraction of photons that are transmitted through the IGM during the EoR is given by (e.g. Mesinger et al. 2014)
\begin{align}
\label{eq:tauigm}
\mathcal{T}_{\rm IGM} = \int_{-\infty}^{\infty}d\Delta v\hs J_{\alpha}(\Delta v) \exp [-\tau_{\rm IGM}(z_{\rm g},\Delta v)], \\ \nonumber
\tau_{\rm IGM}(z_{\rm g},\Delta v)=\taud(z_{\rm g},\Delta v)+\tauHII(z_{\rm g},\Delta v)
\end{align} where $J_{\alpha}(\Delta v)$ denotes the line profile of Ly$\alpha$ photons as they escape from the galaxy, and $\taud(z_{\rm g},\Delta v)$ is the given by Eq~\ref{eq:morepatchII}. Eq~\ref{eq:tauigm} shows explicitly that there are two components to the IGM opacity: ({\it i}) $\taud(z_{\rm g},\Delta v)$ describes the opacity in diffuse neutral intergalactic patches, and this component is thus unique to the reionization epoch, and ({\it ii}) $\tauHII(z_{\rm g},\Delta v)$ the opacity of the ionized component of the IGM (which can be substantial, see \S~\ref{sec:IGM}).

To compute $\taud(z_{\rm g},\Delta v)$ we need to know the spatial distribution of neutral intergalactic patches to compute for a given sightline towards a galaxy. These distributions can be obtained from large-scale radiation transfer simulations, or `semi-numerical' simulations (an example of the large scale ionization field was shown in Figure~\ref{fig:reionization}). We also need to know the location of Ly$\alpha$ emitting galaxies within this simulation. A common assumption is that the Ly$\alpha$ emitting galaxies which are luminous enough to be detectable at these high redshifts with existing telescopes, reside in more massive dark matter halos that existed at the time ($M\sim 10^{10}-10^{11}$ $M_{\odot}$). 

Figure~\ref{fig:TIGM} shows examples of probability distributions ({\it left panel}: differential, {\it right panel}: cumulative) for $\mathcal{T}_{\rm IGM}$ for a model for the IGM whose volume averaged neutral fraction is $\langle x_{\rm HI} \rangle =91\%$ at $z=8.6$ \citep[taken from][this redshift was motivated by the claimed detection of a Ly$\alpha$ emission line at this redshift by Lehnert et al. 2010]{D11}. The {\it blue dotted line} represents a model in which Ly$\alpha$ photons escape from galaxies with a Gaussian emission line [$J_{\alpha} (\Delta v)$] centered on their systemic velocities. The {\it red dashed line} \& {\it black solid line} show models in which the Ly$\alpha$ photons emerge from galaxies with redshifted Ly$\alpha$ line as a result from scattering through an optically thick outflow (see \S~\ref{sec:exp}). This redshifted spectral line was modelled by having the photons scatter through a spherical shell model with a column density of $N_{\rm HI}=10^{20}$ cm$^{-2}$, and outflow velocity of $v_{\rm out}=25$ km s$^{-1}$ and $v_{\rm out}=200$ km s$^{-1}$. This plot shows that inhomogeneous reionization allows $>10\%$ of Ly$\alpha$ photons to be transmitted through a highly neutral IGM (which would correspond to an `effective' optical depth $\langle \tau_{\rm GP} \rangle<$2.3) for a non-negligible fraction of galaxies, and that this fraction goes up dramatically when winds affect the Ly$\alpha$ photons on interstellar scales.

One can construct similar PDFs for other (volume averaged) neutral fractions ($\langle x_{\rm HI} \rangle$), and redshifts ($z$), and estimate how much evolution in $\langle x_{\rm HI} \rangle$ is required to explain the observed reduction in the Ly$\alpha$ flux from galaxies at $z>6$. For this calculation, we must therefore assume that the observed reduction is entirely due to a changing $\mathcal{T}_{\rm IGM}$. We stress that radiative transfer in the IGM can (strongly) affect Ly$\alpha$ photons when reionization is completed (see \S~\ref{sec:IGM}), but the question we address here is how much {\it additional} opacity in the IGM is needed in neutral patches to explain the reduction in Ly$\alpha$ flux at $z>6$. For simplicity, we can therefore assume\footnote{This assumption is not only convenient. If the Ly$\alpha$ line profile is processed by scattering through an outflow, then the IGM transmission through the post-reionization IGM can in fact be close to unity. It is also worth pointing out that \citet{Jensen} adopted different initial line profiles such that $\mathcal{T}_{\rm IGM}(z=6)\sim 0.5$, and obtained the same required change in the neutral fraction. This emphasises that these constraints are sensitive to the {\it change} in the IGM opacity at $z>6$, not the overall normalisation.} that $\mathcal{T}_{\rm IGM}(z=6)=1$. We further assume that the EW-PDF of the Ly$\alpha$ line in drop-out selected galaxies at $z=6$ [$P_6({\rm EW})$] is given by an exponential function, i.e. $P_6({\rm EW})\propto \exp(-{\rm EW}/{\rm EW}_{\rm c})$ (where the scale length EW$_{\rm c}=50$ was chosen to match the data at $z=6$). Under these assumptions, we can predict the EW-PDF at $z=7$ [$P_7({\rm EW}) $] to be
\begin{equation}
P_7({\rm EW}) \propto \int d\mathcal{T}_{\rm IGM}\hs P_7(\mathcal{T}_{\rm IGM})P_6({\rm EW}/\mathcal{T}_{\rm IGM}).
\end{equation}

\begin{figure}
\vspace{-77mm}
\vbox{\centerline{\epsfig{file=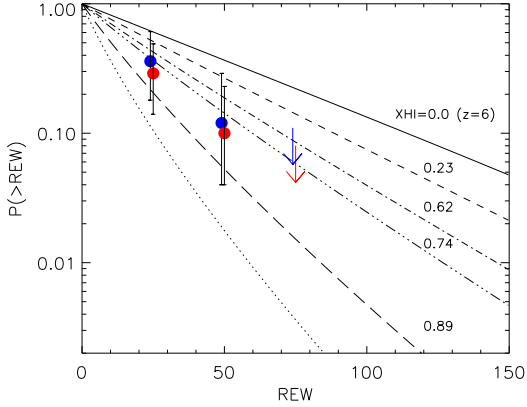,angle=0,width=10.5truecm}}}
\caption[]{Expected evolution of the (cumulative) EW-PDF for a range of HI volume filling factors, $x_{\rm HI}$. The {\it solid line} shows the PDF at $z=6$, which we assume corresponds to $x_{\rm HI}=0.0$. Increasing $x_{\rm HI}$ increases the IGM opacity to Ly$\alpha$ photons, which reduces the PDF. Data points are taken from \citet{P14}. When taken at face value, the data favours  $x_{\rm HI} > 0.5$ at $z=7$ ({\it Credit: from Figure~4 of Pentericci et al. 2014 \textcopyright AAS. Reproduced with permission}).}
\label{fig:Lyafraction2}
\end{figure}

Example PDFs are shown in Figure~\ref{fig:Lyafraction2}. Overplotted are data points at $z=7$ from \citet{P14}. This plot shows that the evolution in the Ly$\alpha$ fraction - when taken at face value -  requires a large neutral fraction at $z=7$, $\langle x_{\rm HI} \rangle \sim 0.5$ is needed to explain the observed drop at $z=7$. What is remarkable about this result, is that the required evolution in $\langle x_{\rm HI} \rangle$ is too fast compared to what is expected based on theoretical models of reionization. 

We note that these results were based on ionization fields generated with semi-numeric models. Similar results have since been obtained by \citet{Jensen}, who used cosmological hydrosimulations in combination with a more complete treatment of radiative transfer of ionizing photons. This agreement is not surprising as semi-numerical methods have been successfully tested against more detailed radiative transfer calculations.  It is worth cautioning that the amount of data on $z=7$ (and even $z<6$) is still small and that, as a consequence, the uncertainties large. Moreover, Taylor \& Lidz (2013) have shown that the inhomogeneity of the reionization process can introduce large variations in the spatial distribution of the Ly$\alpha$ fraction at a fixed $x_{\rm HI}$. They further showed that with the current sample sizes of drop-out galaxies, the observations require $x_{\rm HI} > 0.05$ (95\% CL, although this limit does not include constraints from evolution in the LAE luminosity functions). 
\subsection{Alternatives \& Modeling Uncertainties}
\label{sec:alt}
There are alternative physical explanations that can contribute to reducing the Ly$\alpha$ flux from galaxies at $z>6$ signature. These can be grouped into two categories:\\

In the first category, the IGM opacity evolves rapidly beyond $z=6$, but {\it not} due to the emergence of neutral intergalactic patches. There are several different ways this can happen:
\begin{itemize}[leftmargin=*]
\item The opacity of the ionized intergalactic medium, $\tauHII(z,\Delta v)$, evolves between $z=5.7$ and $z=6.5$. The models of \citet{Laursen11} show that the ionized IGM at $z=6.5$ can transmit $\sim 30\%$ fewer photons than the IGM at $z=5.7$. This is related to the fact that the mean density of the Universe at $z=6.5$ was $\sim 50\%$ higher than at $z=5.7$, which translates to an increase in the Gunn-Peterson optical depth by a factor of $\sim 1.7$ \citep[][]{IGM}. Interestingly, \citet{Ouchi10} have shown that the evolution in the Ly$\alpha$ luminosity functions of LAEs is consistent with a reduction of $\sim 30\%$ in luminosity. It has not been checked whether this model can reproduce the observed drop in the Ly$\alpha$ fraction. It must be noted that these same models predict that the IGM transmits $\mathcal{T}_{\rm IGM}\sim 10-30 \%$ at $z\sim 6$, which may be at tension with observational constraints on the `effective escape fraction' $f_{\rm esc,eff}\equiv \mathcal{T}_{\rm IGM}f_{\rm esc}$ which prefer $f_{\rm esc,eff}\sim 30-50\%$ at $z=6$ \citep{Hayes11,Blanc11,DJ13}.

\item The opacity in self-shielding Lyman Limit Systems (LLSs) explains the observed evolution. At the end of the EoR, the ionising background may have been substantially smaller than post-reionization, as a result of the reduced mean free path of ionizing photons. A reduction in the ionizing background reduces the density of gas that can self-shield against it, and therefore increases the number of systems that can self-shield (with $N_{\rm HI}\gsim 10^{17}$ cm$^{-2}$), which further reduces the mean free path of ionizing photons. This in turn reduces the ionizing background further etc. The number density of absorption systems with $N_{\rm HI} \gsim 10^{17}$ cm$^{-2}$ - the LLSs - in our Universe that can self-shield against an ionizing background therefore depends sensitively on the value of the UV (or ionizing) background.  \citet{Bolton} have shown that LLSs can provide a significant source of opacity to Ly$\alpha$ photons emitted by galaxies, and that the evolution on the number density of LLSs alone can explain the observed drop in the Ly$\alpha$ fraction. The reduced ionising background in these models, still give rise to a significant change in the volume averaged neutral fraction. However, the change required to explain the observed reduction in Ly$\alpha$ flux corresponds only to $x_{\rm HI} \sim 0.1-0.2$. More recently, \citet{Mesinger14} have shown however that the abundance of LLS may have been overestimated, and they recover the large required change of $\Delta x_{\rm HI} \sim 0.5$.

\item Finally, \citet{Finkcolors} speculate that it may be related to the ratio of the gas accretion rate onto galaxies and their star formation rate, which has been inferred observationally to increase by $\sim 40\%$ from $z=6$ to $z=7$ \citep{Papovich11}. Enhanced gas accretion onto galaxies at $z=7$ compared to $z=6$ at a fixed star formation rate -and hence intrinsic Ly$\alpha$ luminosity - can have two effects: ({\it i}) it can increase the opacity of the infalling circum-galactic medium between $z=6$ and $z=7$ (like the first point discussed above), and ({\it ii}) it can cause the line profile emerging from galaxies at $z=7$, $J_{\alpha}(\nu)$, to contain a larger fraction of blue-shifted photons as a result of scattering through a more prominent inflowing component. This enhanced fraction of flux on the blue side of the line would make the Ly$\alpha$ flux more subject to scattering in the ionized IGM (see \S~\ref{sec:IGM}), and hence again enhance its opacity. This possibility has not been investigated quantitatively yet.

\end{itemize}

In the second category, the intrinsic Ly$\alpha$ luminosity (i.e. prior to scattering through the IGM) decreases at $z>6$. Two processes have been considered:
\begin{itemize}[leftmargin=*]

\item The fraction of photons that escape from galaxies decreases at $z>6$. \citet{DF12} argued that dust could mimick a reionization signature. This requires a non-trivial impact of the dust on Ly$\alpha$: As mentioned in \S~\ref{sec:ISM} $\fesca$ is anti-correlated with the dust content of galaxies. The inferred increase $\fesca$ (or more accurately $\fesce$) towards higher redshift can be connected to the reduced dust content of galaxies towards higher redshifts, as inferred from their reddening \citep[e.g.][]{Finkcolors,Bouwenscolor}. The dust content of galaxies keeps decreasing at $z>6$. For this reason, \citet{DF12} invoked an evolution in the dust {\it distribution}. In particular, they require a clumpy dust distribution to boost the Ly$\alpha$ escape fraction at $z=6$ (see \S~\ref{sec:dust}), and not at $z=7$. We note that in this model the required boost in the Ly$\alpha$ EW at $z=6$ is $\sim 1.5$. Recent radiative transfer models through clumpy, dusty media have found such boosts to be rare \citep[][see \S~\ref{sec:dust}]{Laursen13}. 

\item The intrinsic Ly$\alpha$ emission is reduced at $z>6$. Ly$\alpha$ emission is powered by recombination following photoionization inside HII regions, and the Ly$\alpha$ luminosity of a galaxy, $L_{\alpha} \propto (1-\fesc)$, where $\fesc$ denotes the escape fraction of ionizing photons.  Direct observational constraints on the escape fraction, and its redshift dependence are still highly uncertain. However, there are several lines of indirect evidence that $\fesc$ increases with redshift. Measurements of the redshift dependence of the photoionization rate of the Ly$\alpha$ forest in combination with the observed redshift evolution of the UV-LF of drop-out galaxies, suggest that $\fesc$ increases quite rapidly with redshift at $z \gsim 4$ \citep[e.g.][]{Kuhlen,Mitra13}. \citet{Dijkstra14} found that models that incorporate this redshift evolution of $\fesc$ reduce the required global neutral fraction of $\Delta x_{\rm HI}\sim0.2$. Moreover, ionizing photons escape efficiently along paths with $N_{\rm HI} \lsim 10^{17}$ cm$^{-2}$. These same paths enable Ly$\alpha$ photons to escape close to line centre \citep{Behrens14,Verhamme14}. If $\fesc$ indeed increases from $z=6\rightarrow 7$, then this may cause the average Ly$\alpha$ spectral line shape to shift closer to line-center. This in turn could increase the opacity of the ionized gas in the IGM ($\tauHII(z,\Delta v)$, see \S~\ref{sec:IGM}), and further reduce the required $\Delta x_{\rm HI}$.

\end{itemize}

\section{Summary \& Outlook}
\label{sec:outlook}

I have provided a pedagogical review of why and how Ly$\alpha$ emitting galaxies provide a useful probe of the Epoch of Reionization (EoR), and what existing observations of these galaxies tell us on the EoR: Existing observations indicate that Ly$\alpha$ emission is `suddenly' suppressed at $z>6$, which contrasts with the observed redshift evolution of the visibility of the Ly$\alpha$ line at lower redshifts (\S~\ref{sec:obs}). If we wish to attribute the observed reduction in Ly$\alpha$ flux at $z>6$ to the emergence of neutral intergalactic patches, then this require a rapid evolution in the volume-averaged neutral fraction of Hydrogen, $\Delta x_{\rm HI} \sim 0.5$ (\S~\ref{sec:interpretation}). I discussed various alternative explanations for this observation in \S~\ref{sec:alt}.

Current uncertainties on the data are large, mostly due to the limited number of Ly$\alpha$ emitting galaxies at $z>6$, and makes it difficult to distinguish between different models. An additional important uncertainty is the Ly$\alpha$-EW PDF at $z=6$ (and $z<6$). We expect this quantity to be much better constrained in the near future via measurements with e.g. MUSE\footnote{http://www.eso.org/sci/facilities/develop/instruments/muse.html} on VLT, with a quoted line flux sensitivity of $F_{{\rm Ly}\alpha}\sim 4 \times 10^{-19}$ erg s$^{-1}$ cm$^{-2}$ (in Wide Field Mode, over 80hrs of observation), which corresponds to $L_{\alpha}\sim 1.6\times 10^{41}$ erg s$^{-1}$ at $z=6$.

Constraints on the Ly$\alpha$ flux from galaxies at $z>6$ have been (and will be) improving gradually via spectroscopic follow-up of high-z drop-out galaxies. These measurements are difficult and typically require $\sim 10$ hr per galaxy on 8m telescopes. It is worth emphasising that non-detections of the Ly$\alpha$ line can provide extremely useful constraints on reionization \citep[e.g.][]{Treu12}. The Keck Cosmic Web Imager\footnote{http://www.srl.caltech.edu/sal/keckcosmic-web-imager.html} is designed to perform high-precision spectroscopy on faint objects including Ly$\alpha$ emitting galaxies at $5 \lsim z \lsim 7$ \citep[e.g][]{Martin10}, and is expected to significantly improve our knowledge of the evolution of the Ly$\alpha$-emission from galaxies at these redshifts. Complementary constraints can be provided by spectroscopic observations of intrinsically faint, gravitationally lensed galaxies. For example, the Grism Lens-Amplified Survey from Space (GLASS) consists of grism spectroscopy of the core and infall regions of 10 galaxy clusters to look for line emission from gravitationally lensed high-redshift galaxies \citep{Schmidt14}. Finally, Hyper Suprime-Cam\footnote{http://www.naoj.org/Projects/HSC/} on the Subaru telescope has a much larger field-of-view than the existing Suprime-Cam. With this camera, Subaru will be able to increase the sample of LAEs at $z=6.5$, $z=7.0$, and $z=7.3$ by one (possibly two) orders of magnitude, which is enough to detect a reionization-induced signature on the clustering of these LAEs \citep[][]{Jensen}.

Our understanding of the precise impact of reionization on the visibility of the Ly$\alpha$ line, is still limited by our knowledge of radiative transfer process on interstellar (and therefore intergalactic) scales. The employment of near IR spectrographs allows us to measure (rest frame) optical nebular lines such as $H\alpha$, [OII], and [OIII], which allow for better constraints on (and therefore understanding of) the radiative transfer processes.  Recently, the first simultaneous measurements of the Ly$\alpha$ spectral line shape and other rest-frame optical nebular emisson lines have been reported (see the discussion in \S~\ref{sec:ISM}). It has already been shown that these lines provide much stronger constraints on radiative transfer models, than observations of only the Ly$\alpha$ line. For example, these observations can determine the overall shift of the Ly$\alpha$ with respect to a galaxy's systemic velocity, which provides a clean indications of whether outflows are present or not (see \S~\ref{sec:ISM}). 

Furthermore, the Hobby Eberly Telescope Dark Energy Experiment \citep[HETDEX][]{Hill08}\footnote{http://www.hetdex.org} aims to study dark energy by measuring the clustering properties of 0.8 million Ly$\alpha$ selected galaxies at $z=1.5-3.8$. Measurements of the clustering, and luminosity functions will allow for much better constraints on radiative transfer processes on interstellar \& circum galactic levels \citep[e.g.][]{Zheng11,WD11,Behrens13}. We therefore expect the prospects for making progress on the issues that we are currently facing to be very promising.\\

{\bf Acknowledgments} I thank: COSPAR for travel funds as this review grew out of a review talk given at COSPAR in Mysore in 2012; Andrei Mesinger for kindly providing Figure~\ref{fig:reionization} \& Max Gronke for providing Figure~\ref{fig:redist}; Benedetta Ciardi, Luke Barnes, Peter Laursen,  Garrelt Mellema, and an anonymous referee for careful reading of the manuscript and for many helpful comments. I thank my colleagues for their permission to use their figures for this review.

%\end{multicols}

\end{document}